\documentclass[a4paper,11pt]{article}
\pdfoutput=1

\usepackage{jcappub}

\ifx\pdfoutput\undefined
\usepackage[dvips,bookmarks=false]{hyperref}	
\else
\usepackage{hyperref}	
\fi
\hypersetup{colorlinks,bookmarksopen,bookmarksnumbered,citecolor=blue,
linkcolor=blue,pdfstartview=FitH,urlcolor=blue}

\usepackage{amssymb} 
\usepackage{graphicx}
\usepackage{amsmath}
\usepackage{xcolor}
\usepackage{physics}
\usepackage{mathtools}
\usepackage[normalem]{ulem}

\def\beq{\begin{equation}}
\def\eeq{\end{equation}}
\def\beqa{\begin{eqnarray}}
\def\eeqa{\end{eqnarray}}

\newcommand{\N}{\mathcal{N}}
\newcommand{\nl}{\nonumber \\}

\renewcommand{\r}{\vec{r}}
\renewcommand{\v}{\vec{v}}
\newcommand*{\sinc}{\mathrm{sinc}}

\title{A self-consistent wave description of axion miniclusters and their survival in the galaxy}

\author[a]{Virgile Dandoy,}
\author[a]{Thomas Schwetz,}
\author[a,b,c]{Elisa Todarello}

\affiliation[a]{Institut f\"ur Astroteilchenphysik, Karlsruhe Institute of Technology (KIT),\\ 
Hermann-von-Helmholtz-Platz 1, 76344 Eggenstein-Leopoldshafen, Germany}

\affiliation[b]{Dipartimento di Fisica, Universit\`a di Torino, via P. Giuria 1, I–10125 Torino, Italy}
\affiliation[c]{Istituto Nazionale di Fisica Nucleare, Sezione di Torino, via P. Giuria 1, I–10125 Torino, Italy}

\emailAdd{virgile.dandoy@kit.edu}
\emailAdd{schwetz@kit.edu}
\emailAdd{elisamaria.todarello@unito.it}

\abstract{
We present a solution of the Schr\"odinger--Poisson system based on the WKB ansatz for the wave function. In this way we obtain a description of a gravitationally bound clump of axion dark matter by a superposition of energy eigenstates with random phases. It can be applied to any self-consistent pair of radial density distribution and phase space density $f(E)$ related by Eddington's formula. We adopt this as a model for axion miniclusters in our galaxy and use it to study the mass loss due to a star encounter by using standard perturbation theory methods known from quantum mechanics. Finally, we perform a Monte Carlo study to estimate the surviving fraction of axion miniclusters in the dark matter halo of our galaxy. We find that the reaction to perturbations and the survival probability depend crucially on the density profile. Weakly bound clusters are heated up and eventually destroyed, whereas more strongly bound systems get even more compact as a result of perturbations and are driven towards an axion star configuration. 
}

\begin{document}
\maketitle
\flushbottom

\section{Introduction}

The QCD axion \cite{Weinberg:1977ma, Wilczek:1977pj,Peccei:1977ur, Peccei:1977hh} is one of the best motivated dark matter candidates \cite{Preskill:1982cy,Abbott:1982af, Dine:1982ah}, offering a well-defined particle physics embedding as well as a generic way to produce a relic dark matter abundance in the early Universe, see \cite{Sikivie:2006ni, Marsh:2015xka} for reviews of axion cosmology. A generic feature of the so-called post-inflationary scenario is the appearance of axion miniclusters. These are gravitationally bound clumps of axion dark matter, with typical masses and sizes determined by the Hubble volume at the time when the axion starts to oscillate. This has been first pointed out in the seminal paper by Hogan and Rees \cite{Hogan:1988mp}, and subsequently studied by a number of authors, see Refs.~\cite{Kolb:1993zz,Kolb:1993hw,Zurek:2006sy,Hardy:2016mns,Enander:2017ogx,Vaquero:2018tib,Buschmann:2019icd,Eggemeier:2019khm,Ellis:2020gtq,Xiao:2021nkb} for an incomplete list. An important question in this context is the abundance and survival probability of
axion miniclusters in the galaxy up to today. Their presence would have
dramatic consequences for direct axion dark matter searches (see e.g.,
\cite{ADMX:2019uok,ADMX:2021nhd,MADMAX:2019pub}), and may offer
additional signatures, such as radio signals
\cite{Tkachev:2014dpa,Hook:2018iia,Edwards:2020afl} or gravitational lensing
\cite{Kolb:1995bu,Fairbairn:2017dmf,Fairbairn:2017sil,Katz:2018zrn,Dai:2019lud,Ellis:2022grh}.
First studies of the stability of axion miniclusters under encounters with nearby stars in the galaxy or other tidal effects have been performed in Refs.~\cite{Tinyakov:2015cgg,Dokuchaev:2017psd,Kavanagh:2020gcy}.

The generic framework to describe axion miniclusters is the
Schr\"odinger-Poisson system. Due to the extremely high occupation
numbers, the axion energy density can be described by a classical
field, governed by an equation equivalent to the Schr\"odinger
equation, where the potential in turn is determined by the Poisson
equation. In this work we develop a wave description of axion
minicluster in terms of stationary solutions of the Schr\"odinger
equation in the WKB approximation. Adopting a random phase ansatz, we
can build a self-consistent solution of the Schr\"odinger-Poisson
system by considering a superposition of energy eigenstates with the
appropriate weights, the latter being determined by a distribution
function $f(E)$. Our approach is similar to the one first developed by
Kaiser and Widrow \cite{Widrow:1993qq} in a cosmological context, but
adapted to the minicluster problem. For further discussions and
applications of the random phase ansatz see, for instance,
Refs.~\cite{Foster:2017hbq,Lin:2018whl,Knirck:2018knd,Hui:2020hbq}. Our
approach allows a conceptually consistent description of miniclusters
in terms of a classical field. Moreover, the WKB 
approximation offers a well-defined framework to justify the
transition to a description of miniclusters in terms of particles.

In the second part of our work, we study the stability of minicluster under tidal disruption by star encounters. Within our framework, we can apply standard quantum mechanics methods to calculate transitions between energy levels due to a perturbation. We compare our results to a similar study performed by Kavanagh {\it et al.},~\cite{Kavanagh:2020gcy} in the particle picture. Although the field and particle descriptions lead to rather different density profiles after the encounter with a star, we find that the global effect on the total mass and radius after re-virialization of the system is very similar in both formalisms.

We study the stability against tidal effects for two representative examples for the density profile of the axion minicluster, namely the so-called Lane-Emden and the Hernquist profiles. The first one is an example of a relatively loosely bound system, whereas the latter corresponds to a more strongly bound configuration and is more similar to the NFW profile motivated by N-body simulations. We find substantially different behaviours for these two cases with respect to tidal perturbations. Finally, we perform Monte-Carlo simulations based on a semi-realistic distribution of stars inside the Milky Way, to determine the survival probability of axion miniclusters as a function of their mass and size.

The outline of the remainder of this paper is as follows. In section~\ref{sec:axion-clump} we discuss the wave descriptions of axion miniclusters based on the WKB approximation and provide the specific realisations for the Lane-Emden and Hernquist profiles. In section~\ref{sec:perturbation} we consider the perturbation of the minicluster by tidal forces in a stellar encounter. We calculate the transitions between energy levels due to the perturbation and consider the modified minicluster properties after re-virialization after the encounter. In section~\ref{sec:survival} we discuss the survival probability of axion miniclusters in the solar neighbourhood by taking into account the distributions of stars in our galaxy as well as orbital parameters of the minicluster population. We find that the results depend sensitively on the assumed density profile for the miniclusters. We conclude and provide a discussion of our results in section~\ref{sec:conclusion}. Supplementary material is provided in appendices~\ref{app:mapping}, \ref{app:virial}, \ref{app:coefficients}, \ref{app:distribution}. 

For numerical calculations we will adopt always an axion mass of $m_a = 10^{-5}$~eV. Typical scales for axion minicluster mass and radius are $M\sim 10^{-13}\,M_\odot$ and $R \sim 10^{-6}$~pc. For convenience, we note that $10^{-13}\,M_\odot \approx 2\times 10^{17} \, {\rm kg} \approx 2.7\times 10^{-6} M_{\rm moon}$ and
$10^{-6} \, {\rm pc} \approx 0.2\, {\rm AU} \approx 3\times 10^7\,{\rm km}$.

\section{Self-consistent axion clump in the wave description}
\label{sec:axion-clump}

In this section, we construct a wave function for a self-gravitating clump
of axions and discuss its properties. Note that the term ``wave function'' refers to the classical field for the non-relativistic axion.
We start by discussing the WKB approximation in Subsec.~\ref{sec:formalism}, then we build a self-consistent solution from superpositions of WKB energy eigenstates in Subsec.~\ref{sec:self-cons}. In Subsec.~\ref{sec:LE} we introduce two specific examples for the density profile, the so-called Lane-Emeden and Hernquist profiles, which we are going to use to study the impact of stellar encounters in the following.

\subsection{WKB approximation}
\label{sec:formalism}

Let us consider stationary solutions, i.e., the axion wave function will be in an energy eigenstate. Assuming non-relativistic axions, the relevant wave equation is
\beq\label{eq:Schroedinger}
H_0\psi_E(\r\,) \equiv
\left[-\frac{1}{2m_a}\nabla^2 + m_a\phi(\r\,) \right]\psi_E(\r\,) = E \psi_E(\r\,) \enspace,
\eeq
where $m_a$ is the axion mass and $\phi(\r\,)$ is the (dimensionless) gravitational potential. We assume now a spherical symmetric potential $\phi(r)$. Then we follow the usual procedure to separate the wave function in radial and angular components and consider angular momentum eigenstates with quantum numbers $l,m$ and energy eigenstates with eigenvalues $E_n$:
\beq\label{eq:psi_decomp}
\psi_{nlm}(\r\,) = R_{nl}(r)Y_{lm}(\theta,\varphi)\enspace,
\eeq
with the spherical harmonics $Y_{lm}(\theta,\varphi)$.
The radial wave functions $R_{nl}(r)$ (with $r=|\r\,|$) are normalized such that
%
%
\beq\label{eq:normR}
\int_0^\infty dr\,r^2\, R_{nl}R^*_{n'l} = \delta_{nn'} \enspace.
\eeq
Now we insert the ansatz Eq.~\eqref{eq:psi_decomp} in Eq.~\eqref{eq:Schroedinger} and obtain a wave equation for $R_{nl}(r)$ with the effective potential 
\beq\label{eq:Veff}
V_l(r) = \frac{l^2}{2m_a r^2}+m_a \phi(r) \enspace,
\eeq
where we assumed $l\gg 1$; we will show later that this limit is a consequence of the WKB approximation.\footnote{For the WKB approximation of spherical potentials one should replace $l(l+1)\to (l+1/2)^2$ in the effective potential \cite{Langer:1937qr}. In the limit $l\gg 1$ we have $(l+1/2)^2 \approx l(l+1) \approx l^2$.} The wave equation for $R_{nl}(r)$ is solved
under the WKB approximation (sometimes called semi-classical approximation), i.e., assuming that the effective potential $V_l(r)$ 
is varying slowly compared to the deBroglie wave-length, see e.g., \cite{sakurai}:
\beq\label{eq:WKB}
\frac{\lambda}{2\pi} \equiv \frac{1}{\sqrt{2m_a (E_n - V_l)}} \ll
\frac{E_n-V_l}{ \left| dV_l / dr \right|}  \sim D\,,
\eeq
where $D$ is the characteristic size of the system. At next-to-leading order in this approximation the solution for $R_{nl}(r)$ is given by
\footnote{The solution above obviously does not hold in the neighbourhood of the classical turning points. For our numerical computation in the following, we will replace Eq.~\eqref{eq:R}, with the Langer functions discussed in Refs.~\cite{Langer:1937qr,Dingle:1956}. These functions reduce to the WKB solution far from the turning points and to the appropriate Airy function in the neighbourhood of the turning points.}
\beq\label{eq:R}
R_{nl}(r) = \frac{1}{\sqrt{\N_{nl}}}\,\frac{1}{r}\,\frac{1}{\left[2m_a (E_n - V_l(r))\right]^{1/4}}\ 
\sin\left(\int^r dr'  \sqrt{2m_a (E_n - V_l(r'))} + c\right)\enspace.
\eeq
For bound states with $E_n < 0$ we obtain for the normalization constant $\N_{nl}$ after averaging over fast oscillations
\beq
\N_{nl} = \frac{1}{2} \int_{r_1(n,l)}^{r_2(n,l)}dr\,\frac{1}{\sqrt{2m_a (E_n - V_l(r))}} 
\,,\qquad c = \frac{\pi}{4}\,,\qquad (E_n < 0) \enspace, \label{eq:N}
\eeq
where $r_{1,2}(n,l)$ are the classical turning points defined by $E_n = V_l(r)$, and we assume that the part of the wave function in the classically forbidden region is negligible. For unbound states, $E > 0$ and the spectrum is continuous. For convenience 
we still keep the index $n$ to label energies also in the continuous case.
To calculate the normalization for $E_n > 0$ we neglect the potential term $V_l(r)$, using that the integral in Eq.~\eqref{eq:normR} is dominated by regions where $|V_l(r)| \ll E_n$. In this approximation we find
\beq
\N_{nl} = \frac{\pi}{2m_a\, dE_n}\,,\qquad c = 0 \,,\qquad (E_n > 0)\enspace.
\eeq
Here we have introduced the differential energy $dE_n$ to ensure that the wave function $R_{nl}$ has the same dimension for positive and negative energies. 
The normalization condition Eq.~\eqref{eq:normR} holds in both cases, with the identification $\delta_{nn'} \leftrightarrow dE_n \delta(E_n - E_{n'})$.

\subsection{Construction of a self-gravitating system}
\label{sec:self-cons}

Our goal is now to use the solutions of the wave equation, Eqs.~\eqref{eq:psi_decomp} and \eqref{eq:R} to construct a self-consistent wave function such that the gravitational potential is generated by the mass distribution itself. To this aim we consider a linear superposition of bound states with $E_n < 0$:
\beq\label{eq:psi_expansion}
\psi(\r,t) = \sum_{nlm} C_{nlm} R_{nl}(r)Y_{lm}(\theta,\varphi) e^{-iE_n t}\enspace.
\eeq
We must now choose the coefficients $C_{nlm}$ such that the potential $\phi(r)$ and the 
axion mass density $\rho(r)$ are related via the Poisson equation:
\beq
\nabla^2 \phi(r) = 4\pi G \rho(r)\enspace,\label{eq:Poisson}
\eeq
with $\rho(r) = m_a \langle |\psi(\r,t)|^2 \rangle$. The meaning of the average is discussed in the following. Assuming that the phases of the different eigenstates are uncorrelated, we adopt the ansatz
\beq
C_{nlm} = \sqrt{(2\pi)^3 f(E_n)\, g_l(E_n) \,dE_n}\ e^{i\phi_{nlm}}\enspace.\label{eq:coeff}
\eeq
Here $\phi_{nlm}$ are random phases drawn from a uniform distribution, the function $f(E)$ plays the role of a phase space density, whereas $g_l(E_n)$ denotes the density of states obtained in the continuum limit: $g_l(E) = dn/dE$, such that $\sum_n \to \int dn \to \int (dn/dE) dE$. The density of states can be obtained by using the WKB quantization condition
\beq
\int_{r_1(n,l)}^{r_2(n,l)}dr \sqrt{2m_a(E-V_l(r))} = \pi \left(n+\frac{1}{2}\right) \,.
\eeq
Considering Eq.~\eqref{eq:N}, it follows that up to a constant factor it is just given by the normalization coefficients: 
\beq
g_l(E) = \frac{2m_a}{\pi}\,\N_{nl}\enspace. \label{eq:g}
\eeq
For large $l$ we can also replace the sum by an integral via the rule $\sum_l \to \int dl$.

For a given axion minicluster, the phases $\phi_{nlm}$ will have some random, but fixed values. In the following we will be interested in an \emph{average} minicluster, by taking the ensemble average of the random phases, such that 
\beq
\langle C_{nlm}C^*_{n'l'm'} \rangle_{ens} = |C_{nlm}|^2 \delta_{nn'}\delta_{ll'}\delta_{mm'} \,.
\eeq
We can calculate now the density profile of the average minicluster as
\beqa
\rho(r) &=& m_a\langle|\psi(\r, t)|^2\rangle_{ens}  \nl
&=& 
m_a\sum_{nlm} |C_{nlm}|^2 R_{nl}^2(r) |Y_{lm}(\theta, \varphi)|^2\nl
&=& 
\frac{4\pi m^2_a}{r^2}\int_{m_a\phi(r)}^0 dE\, f(E)
\int_0^{l_{\rm max}(E, r)} dl \,\frac{l}{\sqrt{2m_a (E - V_l(r))}}\enspace,\label{density_dE_dl}
\eeqa
where in addition to the ensemble average over the random phases, we have also averaged the fast oscillations in $R_{nl}^2$. Note that the ensemble average leads to a spherically symmetric and time independent distribution, as expected. Conversely, in this case, time and angular averages together have the same effect as the ensemble average. 
For given $r$ and $E$, the maximum allowed angular momentum $l_{\rm max}(E,r)$ is defined by $E=V_{l_{\rm max}}(r)$, 
whereas the integral $dE$ runs from the value of the effective potential with $l=0$ to 0.
Using Eq.~\eqref{eq:Veff}, the $l$-integral can be performed and we obtain
\beqa\label{eq:rho}
\rho(r) = 4\pi m_a^2\int_{m_a\phi(r)}^0 dE\, f(E)\, \sqrt{2m_a(E - m_a\phi(r))}\enspace.\label{density_dE}
\eeqa
This expression is identical to a corresponding relation for the phase-space density in spherical astrophysical systems, see e.g., \cite{Galactic}, chapter~4. Since $\phi(r)$ is a monotonic function of $r$ we can invert it, eliminate $r$ and consider the density as a function of the potential instead of the radius: $\rho(\phi)$. Then Eq.~\eqref{eq:rho} can be transformed into an Abel integral equation for $f(E)$ with the solution given by Eddington's formula \cite{eddi, Galactic}:
\begin{equation}\label{eq:eddington}
    f(E) = \frac{1}{2\pi^2m_a^2} \frac{d}{dE}
    \int_{E/m_a}^0 d\phi \, \frac{1}{\sqrt{2m_a(m_a\phi - E)}}
    \frac{d\rho}{d\phi} \,.
\end{equation}
Here we adopt the convention that $f(E) \ge 0$ for $E\le 0$, i.e., the states up to $E=0$ are occupied. This defines the zero-point for the energy scale used for the potential. For a given pair of $\rho(r)$ and $\phi(r)$ related by Poisson equation, we can use eq.~\eqref{eq:eddington} to derive the corresponding density $f(E)$. Hence, this procedure allows to construct a self-gravitating axion cluster based on the WKB approximation for the wave functions.

The total mass of the clump is given by 
\begin{align}
 M &= 4\pi \int_0^\infty dr\, r^2\rho(r) \nonumber\\
   &= 16\pi^2m_a^2 \int_0^\infty dr\, r^2 \int_{m_a\phi(r)}^0 dE\, f(E)\, 
   \sqrt{2m_a(E - m_a\phi(r))}\label{eq:M} \\
   &= 16 \pi^3 m_a \int_{E_{\rm min}}^0 dE f(E) \int_0^{l_{\rm max}(E)} dl\, l \, g_l(E) \,, \label{eq:Mdens}
\end{align}
where in the last step we exchanged the order of integration and used the density of states $g_l(E) = dn/dE$, see Eqs.~\eqref{eq:g} and \eqref{eq:N}. $l_{\rm max}(E)$ is defined by the largest $l$ for which the equation $E = V_l(r)$ has a solution for any $r$, and $E_{\rm min}$ is the minimum of the potential. 

The gravitational binding energy of the clump is 
\beqa\label{eq:W}
W = -4\pi G\int_0^\infty dr \, r^2 \frac{\rho(r)M(r)}{r}  \enspace,
\eeqa
where $M(r)$ is the mass contained within radius $r$. 
As we show in appendix~\ref{app:virial}, the ensemble average of a clump based on eqs.~\eqref{eq:psi_expansion}, \eqref{eq:coeff} is in virial equilibrium, such that $W + 2K = 0$, with $K$ denoting the total kinetic energy, and $E_{\rm tot} = W + K = W/2$. \footnote{Note that $E_\mathrm{tot}\neq \langle H_0 \rangle_{ens} = \sum_{nlm} E_n |C_{nlm}|^2$. The latter quantity corresponds to the energy stored in the field if the potential $\phi$ was a \emph{fixed} external potential. In contrast, $E_{\rm tot} = W + K$ corresponds to the energy stored in the self-gravitating system, where the potential is provided by the mass distribution itself (i.e., the potential diminishes when mass is successively removed).}

\bigskip

We add the following comments: There is a one-to-one correspondence between the density $f(E)$ for the wave functions (defined as occupation of the eigenfunction of the wave equation with energy $E$) and a phase space density of classical point particles $f(\vec{r},\vec{v})$, with the identification $E = m_a v^2/2 + m_a\phi(r)$. We provide a brief discussion of this mapping in appendix~\ref{app:mapping}. While this correspondence is illuminating, here we never need to refer to the particle picture and will stick to the wave interpretation throughout our discussion.

There is some similarity of our approach to the seminal work of Kaiser and Widrow \cite{Widrow:1993qq}, who use the random phase ansatz for plane wave scalar fields and establish a correspondence between the smoothed axion field and the particle phase space density with the purpose of substituting $N$-body particle simulations by studying the time evolution of the Schr\"odinger-Poisson system. In our work we concentrate on self-gravitating stationary systems and benefit from the explicit wave solutions in the WKB regime.

From numerical simulations, one expects that a so-called soliton or axion star appears in the center of a minihalo \cite{Schive:2014hza,Levkov:2018kau,Eggemeier:2019jsu,Chen:2020cef}. This corresponds to close-to-ground-state solutions of the Schr\"odinger--Poisson system, for which the WKB approximation does not hold. Therefore, we will limit the application of our description to cases, where the impact of the soliton on the minihalo can be neglected. We will return to this issue in section~\ref{sec:validity} with a more detailed discussion.

\subsection{The Lane-Emden and Hernquist profiles}
\label{sec:LE}

As we will see below, the response of miniclusters to tidal shocks can be qualitatively different, depending on the density profile. The expected shape of the profile is currently not well understood, see \cite{Ellis:2022grh} for a recent simulation.
In the following we will use two specific realizations for a self-consistent potential--density pair, namely (a) the Lane-Emden (LE) profile for a polytrope of index $n=1$, where $\rho\propto \phi^n$, see e.g., \cite{Galactic} for a discussion and (b) the so-called Hernquist (H) profile \cite{Hernquist:1990be}. The potentials and density profiles are given by 
\begin{align}
\phi_{\rm LE}(r) &= \left\{
\begin{array}{ll}
-\frac{GM}{R}\,\sinc\left( \frac{\pi r}{R} \right) \qquad &   r<R \\[2mm]
\frac{GM}{R} - \frac{GM}{r}  &  r> R 
\end{array} 
\right.  \qquad &\phi_{\rm H}(r) &= -\frac{GM}{(r+r_{\rm H})}  \,,
\label{eq:potLE-H} \\
\rho_{\rm LE}(r) &= \frac{\pi}{4}\frac{M}{R^3}\,\sinc\left( \frac{\pi r}{R} \right) \qquad\quad   r<R 
&\rho_{\rm H}(r) &= \frac{M}{2\pi} \frac{r_{\rm H}}{r(r+r_{\rm H})^3}\,.
\label{eq:densityLE-H}
\end{align}
While the LE ansatz is not fully motivated by first principles, it has the advantage that it describes a self-consistent clump of finite size $R$, with $\rho_{\rm LE}(r) = 0$ for $r> R$. The Hernquist profile resembles the $1/r$ behaviour at small radii, similar to the NFW profile \cite{Navarro:1996gj}, but falls off faster at large radii, such that the total mass is finite for $r\to \infty$. Our choice for the zero-point of energy and potential is consistent with the convention in the previous subsection; it implies $\phi_{\rm LE}(r) \to GM/R$ and $\phi_{\rm H}(r) \to 0$ for $r\to \infty$.

\begin{figure}[t]
\centering
  \includegraphics[width=0.95\linewidth]{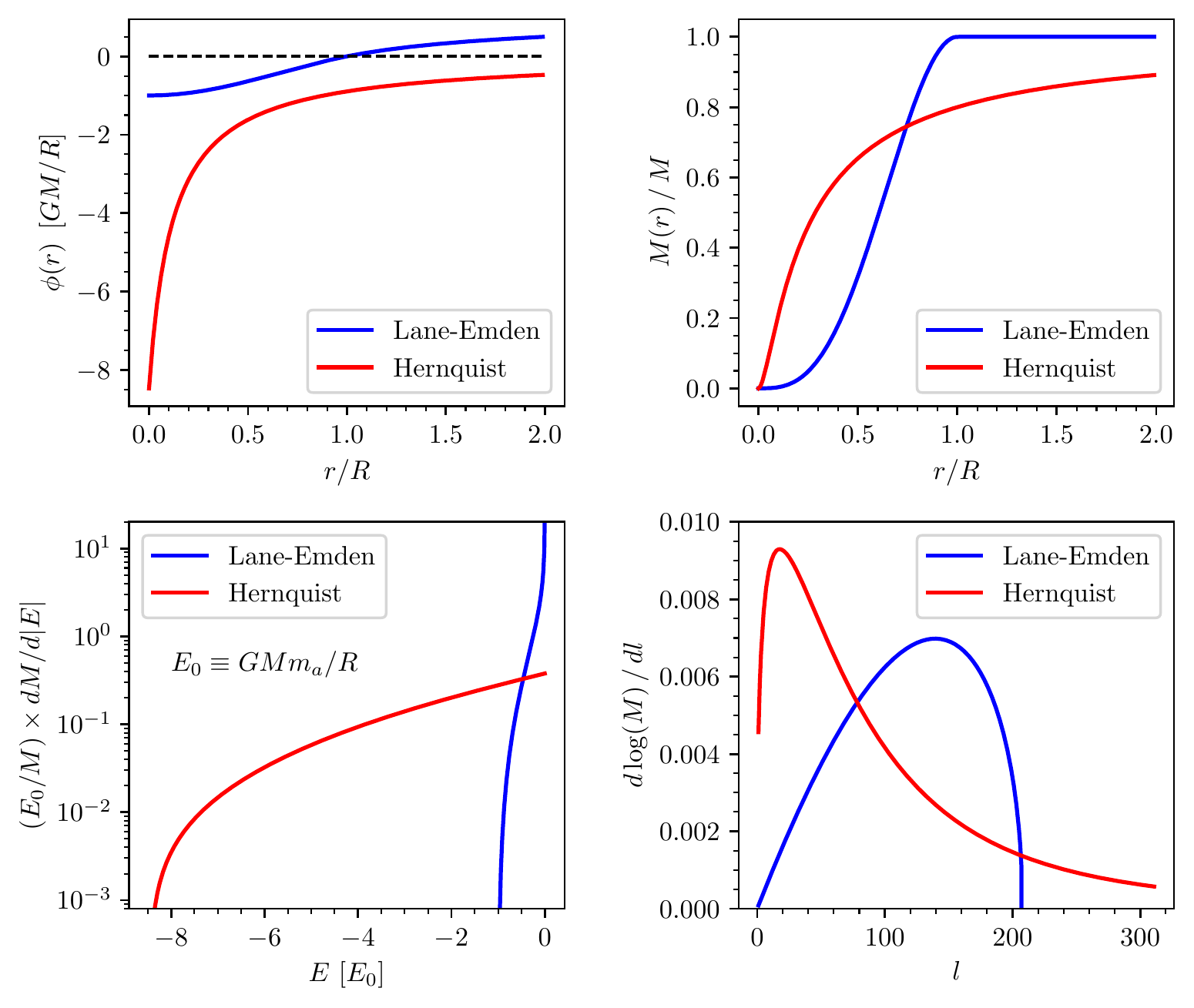}
 \caption{Comparison of the Lane-Emden (blue) and Herquist (red) profiles. We show the potential (top left), the cumulative mass distribution $M(r)$ (top right), and the differential contribution to the total mass of energy states (bottom left) and angular momentum states (bottom right). We take the total mass $M = 10^{-12}M_\odot$, $R = 10^{-6}$~pc, $m_a = 10^{-5}$~eV, 
 and for Hernquist we define $R$ such that 80\% of the total mass is contained in $r<R$, which implies $R = 8.47 \, r_{\rm H}$. The energy unit is $E_0=GMm_a/R$.}
\label{fig:profiles}
\end{figure}

We compare the potential profiles and the cumulative mass distributions for LE and H in the top panels of Fig.~\ref{fig:profiles}. In the figure we have set the Hernquist scale radius $r_{\rm H}$, such that $\beta = 80\%$ of the total mass is contained in $r<R$. For the radius $r_\beta$, which contains a fraction $\beta$ of the total mass, we have
\begin{align}
      r_\beta = r_{\rm H} \, \frac{\beta + \sqrt{\beta}}{1-\beta} \,,
\end{align}
and $r_{0.8} \approx 8.47 \, r_{\rm H}$. In the following we will always adopt the convention of setting $R = r_{0.8}$, when comparing LE and H profiles. We see from the figure that the LE potential is much shallower, with the potential minimum of the Hernquist being a factor $r_\beta/r_{\rm H}$ deeper. For the same mass $M$ and comparable size ($R \simeq r_{0.8}$) the Hernquist potential leads to a much tighter bound core. The gravitational binding energies are easily obtained from  Eq.~\eqref{eq:W}:
\beqa\label{eq:W_LE-H}
W_{\rm LE}  = -\frac{3}{4} \frac{GM^2}{R} \enspace, \qquad
W_{\rm H}  = -\frac{1}{6} \frac{GM^2}{r_{\rm H}} \enspace,
\eeqa
which leads to $W_{\rm H} \approx 1.9 \, W_{\rm LH}$ for $R = 8.47 \, r_{\rm H}$. 

A technical advantage of the LE and H profiles is that the Eddington formula described in the previous section leads to an analytical expression for the phase space density $f(E)$. We have \cite{Binney:1996sv,Hernquist:1990be}
\begin{align}
    f_{\rm LE}(E) =& \frac{1}{m_a^4}\frac{1}{8\sqrt{2}\pi GR^2}\sqrt{-\frac{m_a}{E}}\,, \label{EL_f} \\
    f_{\rm H}(E) =& \frac{1}{m_a^4}\frac{M}{8\sqrt{2}\pi^3 r_{\rm H}^3 v_g^3} 
    \frac{3\sin^{-1}(q) + q\sqrt{1-q^2}(1-2q^2)(8q^4 - 8q^2-3)}{(1-q^2)^{5/2}}  \,,
    \label{H_f}
\end{align}
where
\begin{align}
    q=\sqrt{-\frac{r_{\rm H}}{GM} \frac{E}{m_a}} \,,\qquad 
    v_g = \sqrt{\frac{GM}{r_{\rm H}}} \,,
\end{align}
and $f(E) = 0$ for $E \ge 0$. 

In the bottom left panel of Fig.~\ref{fig:profiles} we show the contribution of each energy level $E$ to the total mass, i.e., $dM/d|E|$, see also Eq.~\eqref{eq:Mdens}. While for LE, only modes with $ER/(GMm_a) > -1$ contribute and the distribution diverges for $E\to 0$, for Hernquist there is a significant contribution from much more strongly bound states. This will have important consequences for the stability against tidal disruption discussed below, which affects mostly loosely bound states close to $E\approx 0$.

In the bottom right panel of Fig.~\ref{fig:profiles} we show the normalized contribution of the angular modes $l$ to the total mass. This can be obtained from Eq.~\eqref{eq:Mdens} by exchanging the order of the $l$ and $E$ integrals. For the LE potential, there is a maximum angular momentum $l_{\rm max}^{\rm LE}$ for which bound states exist. Using Eqs.~\eqref{eq:Veff} and \eqref{eq:potLE-H} one finds:
\begin{align}
    l_{\rm max}^{\rm LE} &\approx 0.607 \, m_a\sqrt{GMR}  \label{eq:lmax1}\\
    &\approx 210 \, \frac{m_a}{10^{-5}\, {\rm eV}} \sqrt{\frac{M}{10^{-12}M_\odot}}
    \sqrt{\frac{R}{10^{-6}\,{\rm pc}}}\,. \label{eq:lmax}
\end{align}
As seen from the figure, the clump will receive contributions from $l$ values ranging from zero up to $l_{\rm max}$ with a distribution peaking roughly around $l_{\rm peak}^{\rm LE}\simeq 0.7 \, l_{\rm max}^{\rm LE}$. Let us now consider the WKB condition Eq.~\eqref{eq:WKB}. Setting $D=R$ and estimating $E-V_l \sim m_aGM/R$, we find that the WKB condition is equivalent to $l_{\rm max} \gg 1$. Hence, in the following we will work in the large $l$ limit, consistent with the assumption that the WKB approximation holds. 

For the Hernquist profile, the shape of the $l$ distribution is rather different, with a peak at significantly lower values, $l_{\rm peak}^{\rm H} \approx 18$ for the parameters chosen in the plot. 
Numerically we find that the peak value has the same scaling with parameters as given in Eq.~\eqref{eq:lmax1} (which follows also from dimensional arguments). 
Note also that for Hernquist there is no maximum $l$ value and the distributions continuously goes to zero for $l\to \infty$.

\section{Perturbation of the clump} \label{sec:perturbation}

We proceed now to calculate the tidal effect on the clump from a star passing by. Assuming the impact parameter $b$ is much larger than the clump radius $R$, we can consider the star to be a point mass. The gravitational potential at the point $\r$ created by the star is
\beq
\phi_*(\r,t) = -\frac{GM_*}{|\r - \r_*(t)|}\enspace.
\eeq
where $\r_*(t)$ is the position of the star at time $t$. To extract the tidal term we expand the denominator in Legendre polynomials. Terms of zeroth and first order describe just effects on the clump as a whole, whereas the leading order tidal effect is provided by the term of order 2:
\beq
\phi_\mathrm{tidal}(\r,t) = -\frac{GM_*}{r_*(t)} \left(\frac{r}{r_*(t)}\right)^2 P_2(\cos\gamma(t))\enspace,
\eeq
where $\gamma$ is the angle between $\r$ and $\r_*(t)$ and $P_2$ denotes the 2nd order Legendre polynomial.
We assume that for the time relevant to the interaction, the star moves in a straight line with velocity $v$ and impact parameter $b$. We then obtain the Hamiltonian of the perturbation
\beq
H_1(\r, t)= m_a\phi_\mathrm{tidal} = -\frac{GM_* m_a r^2}{(b^2+v^2t^2)^{3/2}} P_2(\cos\gamma(t))
= \frac{A r^2}{(1+(\alpha t)^2)^{3/2}} P_2(\cos\gamma(t)) \enspace,
\label{Perturbation}
\eeq
where we have defined
\beq
A=-\frac{GM_*m_a}{b^3}\qquad\qquad \alpha=\frac{v}{b}\enspace.
\eeq

\subsection{Perturbed axion field}

We calculate the effect of $H_1$ on the axion field perturbatively up to second order using standard time-dependent perturbation theory. We will use Dirac notation for convenience, keeping in mind though that our calculation is fully classical.

We expand the axion field as
\beq
\ket{\psi(t)} = \ket{\psi^{(0)}(t)}+\ket{\psi^{(1)}(t)}+\ket{\psi^{(2)}(t)}\enspace.
\eeq
Here $\ket{\psi^{(0)}(t)}$ represent the time dependent unperturbed axion field, whose energy eigenmodes $\ket{nlm}$ and coefficients $C^{(0)}_{nlm}$ were described in the previous section. The corrections satisfy
\beqa
i\partial_t\ket{\psi^{(i)}(t)} &=& H_0 \ket{\psi^{(i)}(t)} + H_1(t) \ket{\psi^{(i-1)}(t)}\enspace,
\eeqa
with $i=1,2$.
We expand onto the energy basis
\beq
\ket{\psi^{(i)}(t)} = \sum_{nlm} C^{(i)}_{nlm}(t)\ e^{-iE t}\ket{nlm}\enspace,
\eeq
and obtain the equations for the coefficients
\beqa
i\partial_t C^{(i)}_{nlm}(t) &=& \sum_{n'l'm'} C^{(i-1)}_{nlm} \bra{nlm}H_1(\r,t)\ket{n'l'm'}\ e^{i(E-E') t}\enspace,
\label{eqci}
\eeqa
where here and in the following we use the short-hand notation $E=E_n$ and $E'=E_{n'}$. 

Setting the initial time $t_0\to -\infty$ and $C^{(1,2)}_{nlm}(-\infty)=0$, we can integrate the equations above for $t\to\infty$.\footnote{When we omit the time argument of $C^{(i)}_{nlm}(t)$ or $\ket{\psi^{(i)}(t)}$ with $i=1,2$, it is implied that the quantity is evaluated at $t\to\infty$.} 
We obtain an expression for the correction to the coefficients 
\beqa
\Delta |C_{nlm}|^2 &\equiv& \ev{|C_{nlm}|^2}_{ens} - |C_{nlm}^{(0)}|^2 \nl
&=&
\ev{|C_{nlm}^{(1)}|^2}_{ens} + 
\ev{{C_{nlm}^{(0)}}^* C_{nlm}^{(2)}
+C_{nlm}^{(0)} {C_{nlm}^{(2)}}^*}_{ens}\enspace.\label{Csquared}
\eeqa
For a derivation and an expression for $\Delta |C_{nlm}|^2$ in the limit $l\gg 1$ see Appendix~\ref{app:coefficients}. We split the variation of the coefficients into two parts, $\Delta^{(+)}|C_{nlm}|^2$ and $\Delta^{(-)}|C_{nlm}|^2$, corresponding to transitions toward and outward the energy level $\ket{nlm}$, respectively:
\beq
\Delta |C_{nlm}|^2 = \Delta^{(-)} |C_{nlm}|^2  + \Delta^{(+)} |C_{nlm}|^2\enspace.
\eeq 
with 
\beqa
\Delta^{(+)}|C_{nlm}|^2 &=& \left(\frac{A}{\alpha} \right)^2
\sum_{n',B}  |C_{n'lm}^{(0)}|^2 \,
\left|{\bra{nl}}r^2\ket{n'l}\right|^2 g_{nn'} \label{eq:D+}\\
\Delta^{(-)}|C_{nlm}|^2 &=& -|C_{nlm}^{(0)}|^2 \left(\frac{A}{\alpha} \right)^2   \sum_{n'}  \left|{\bra{nl}}r^2\ket{n'l}\right|^2 g_{nn'}\enspace, \label{eq:D-}
\eeqa
where 
$|\bra{nl}r^2\ket{n'l}|^2$ is the matrix element from sandwiching the perturbation $H_1$ between the energy eigenstates (c.f., Eqs.~\eqref{Perturbation} and \eqref{eqci}), and $g_{nn'}$ is a window function arising from the time integral of the interaction. Precise definitions and detailed expressions are provided in Appendix~\ref{app:coefficients}. 
Eq.~\eqref{eq:D+} describes transitions towards the state $\ket{nlm}$, and the sum over $n'$ is only over bound states (indicated by $B$), since only these are populated before the interaction, as $C^{(0)}_{n'lm} = 0$ for unbound states. In contrast, Eq.~\eqref{eq:D-} corresponds to transitions $\ket{nlm} \to \ket{n'lm}$ and the sum over $n'$ goes over bound as well as unbound states. The expressions in Eqs.~\eqref{eq:D+} and \eqref{eq:D-} hold after averaging over the $m$ quantum number, see Appendix~\ref{app:coefficients} for details.


%

\subsection{Energy transferred by the star}
\label{sec:transfE}

From the new coefficients, it is possible to get an expression for the energy deposit of the star into the system. We obtain two contributions for the energy variation as
\begin{equation}
\Delta E^{(-)} =\sum_{nlm, B} E_{nl}\,\Delta^{(-)}|C_{nlm}|^2 \,,
\end{equation}
\begin{equation}
\Delta E^{(+)} =\sum_{nlm} E_{nl}\,\Delta^{(+)}|C_{nlm}|^2 \,,
\end{equation}
and the total change in energy is given by the sum of these two contributions. We find
\begin{equation}
\begin{split}
    \Delta \mathcal{E} =16\pi^2m_a  \int_0^{l_{\rm max}} dl \,2l \int_{E_{\rm min}(l)}^0 dE\, & \,f(E)\mathcal{N}_{nl} \int_{E_{\rm min}(l)}^\infty dE'\times \\
    & g_l(E')\,(E'-E)\left( \frac{A}{\alpha}\right)^2 |\bra{nl}r^2\ket{n'l}|^2 g_{nn'} \,,
\end{split}
\label{eq:DE1}
\end{equation}
where $E_{\rm min}(l)$ is the minimum value of the effective potential $V_l(r)$ and
we have used Eqs.~\eqref{eq:D-}, \eqref{eq:D+}, the explicit expression for the coefficients $|C_{nlm}^{(0)}|^2$ from Eq.~\eqref{eq:coeff}, and the density of states, Eq.~\eqref{eq:g}, to transform the last sum over $n'$ into an integral.

To proceed, we adopt now an approximation known as ``impulse approximation'' in the particle picture \cite{Spitzer:1958}. It corresponds to the case that the time scale of the star encounter $\sim b/v$ is much shorter than the dynamical time scale of the bound system $R \sqrt{R/GM}$, such that bound particles can be considered ``at rest'' during the encounter. In fact, we have

\begin{align}
    \frac{b}{R} \, \frac{1}{v} \sqrt{\frac{GM}{R}}
    &\simeq 2\times 10^{-7} \frac{b}{R} 
    \left(\frac{10^{-3}}{v}\right)
    \left(\frac{M}{10^{-12}M_\odot}\right)^{1/2}
    \left(\frac{10^{-6}\, {\rm pc}}{R}\right)^{1/2} \,,\label{impulse}
\end{align}
which shows that for typical parameters the approximation  is  justified.

The validity of this approximation has been well verified in particle simulations and is commonly adopted in studies of minihalo disruptions \cite{Galactic,Green:2006hh,Schneider:2010jr, Dokuchaev:2017psd,Delos:2019tsl,Kavanagh:2020gcy}.
In appendix~\ref{sec:impulsapprox}, we derive Eq.~\eqref{impulse} in the wave formalism. As we discuss also in the appendix, in practical terms it implies that $g_{nn'}$ can be considered as constant, $g_{nn'}\approx 1/2$, and pulled out of the integral over $E'$ in Eq.~\eqref{eq:DE1}. Then the remaining integral can be simplified by using a so-called  generalized quantum mechanical sum rule~\cite{Sanwu,Belloni_2008} \footnote{In order to apply the sum rule, the sum has to be over a complete set of states. As the sum runs over both, bound and unbound states in Eq.~\eqref{eq:sumrule}, this condition is fulfilled. Furthermore, the sum over $l$ and $m$ is trivial, because the operator $r^2$ does not act on the angular part.}
\begin{equation}
\begin{split}
  \int_{E_{\rm min}(l)}^{\infty}dE'\, g(E') (E'-E) |\bra{nl}r^2\ket{n'l}|^2&= \\
  \sum_{n'} (E_{n'} - E_n) |\bra{nl}r^2\ket{n'l}|^2&=
  \frac{2}{m_a}\bra{nl}r^2\ket{nl} \,.
\end{split}
\label{eq:sumrule}
\end{equation}

Under these assumptions, the energy transferred by the star is expressed as 
\begin{equation}
    \Delta \mathcal{E} =16\pi^2m_a  \int_0^{l_{\rm max}} dl \,2l \int_{E_{\rm min}(l)}^0 dE\,  f(E)\mathcal{N}_{nl}\, \delta E(E,l) \,,
    \label{total injected energy}
\end{equation}
where we defined 
\begin{equation}
\begin{split}
     \delta E (E,l)&=  \left(\frac{2GM_*}{b^2v}\right)^2 \, \frac{m_a}{4}\bra{nl}r^2\ket{nl} \,.
\label{injected energy}
\end{split}
\end{equation}
%
This last quantity corresponds to the energy shift which a level $(E,l)$ receives due to the impact of the star. Note that from Eq.~\eqref{injected energy} it follows that $\delta E(E,l)$ is positive, i.e., levels can only be moved up, towards less strongly bound states. Furthermore, $\delta E(E,l)$ is a growing function of $E$ for fixed $l$. 
Interestingly, the expression for the injected energy calculated through our formalism is in very close analogy with the one derived in a particle description, see for instance in Refs.~\cite{Spitzer:1958,Green:2006hh,Delos:2019tsl}. 
In the particle picture the expectation value in Eq.~\eqref{injected energy} is replaced by $r^2$ and the particles therefore receive an energy shift depending on their location. This expectation value is then the natural generalization in the wave formalism.

\subsection{Properties of the clump after the encounter}
As mentioned, according to Eq.~\eqref{total injected energy}, each energy level is shifted as $E\rightarrow E+\delta E(E,l)$. This implies that all energy levels with $|E|<\delta E(E,l)$ will transfer their occupation number to positive energy levels, i.e., to unbound states. We can therefore define for each angular momentum $l$ a critical energy $\Tilde{E}(l)$ by
\begin{align}\label{eq:Etilde}
\Tilde{E}(l) \equiv \delta E (\Tilde{E},l) \,,    
\end{align}
such that all energy levels with $|E|<\Tilde{E}(l)$ will be removed from the clump.
The new properties of the clump, i.e the new density profile, mass and radius after the encounter can then be calculated by removing the ejected states from the clump, as we are going to discuss in the following.


The change in the density profile is obtained as (c.f., Eq.~\eqref{density_dE_dl})\footnote{Although in general $\Tilde{E}(l)$ defined in Eq.~\eqref{eq:Etilde} depends on $l$, we find that numerically this dependence is very weak. In the approximation of neglecting the $l$-dependence, we can exchange the order of the $l$ and $E$ integrals, and the integral over $l$ can be carried out analytically. We have checked that this approximation does not affect significantly the final result. Similar simplifications can be applied to Eqs.~\eqref{eq:EK} and \eqref{eq:Einj} below.}
\begin{equation}
\Delta \rho(r) = 4\pi \frac{m_a^2}{r^2}\int^{l_{\rm max}(r)}_0 dl\,l  \int^0_{-\Tilde{E}(l)} dE\, \frac{f(E)}{\sqrt{2m_a\left(E-V_l(r)\right)}} \,,
\end{equation}
where $l_{\rm max}(r)$ is the maximum value of the angular momentum $l$ at a given location $r$. This implies a change in mass of
\begin{align}
\Delta M = 4\pi \int_0^\infty dr\, r^2\, \Delta\rho(r) \,.    
\end{align}

\begin{figure}[t]
\centering
  \includegraphics[scale=0.32]{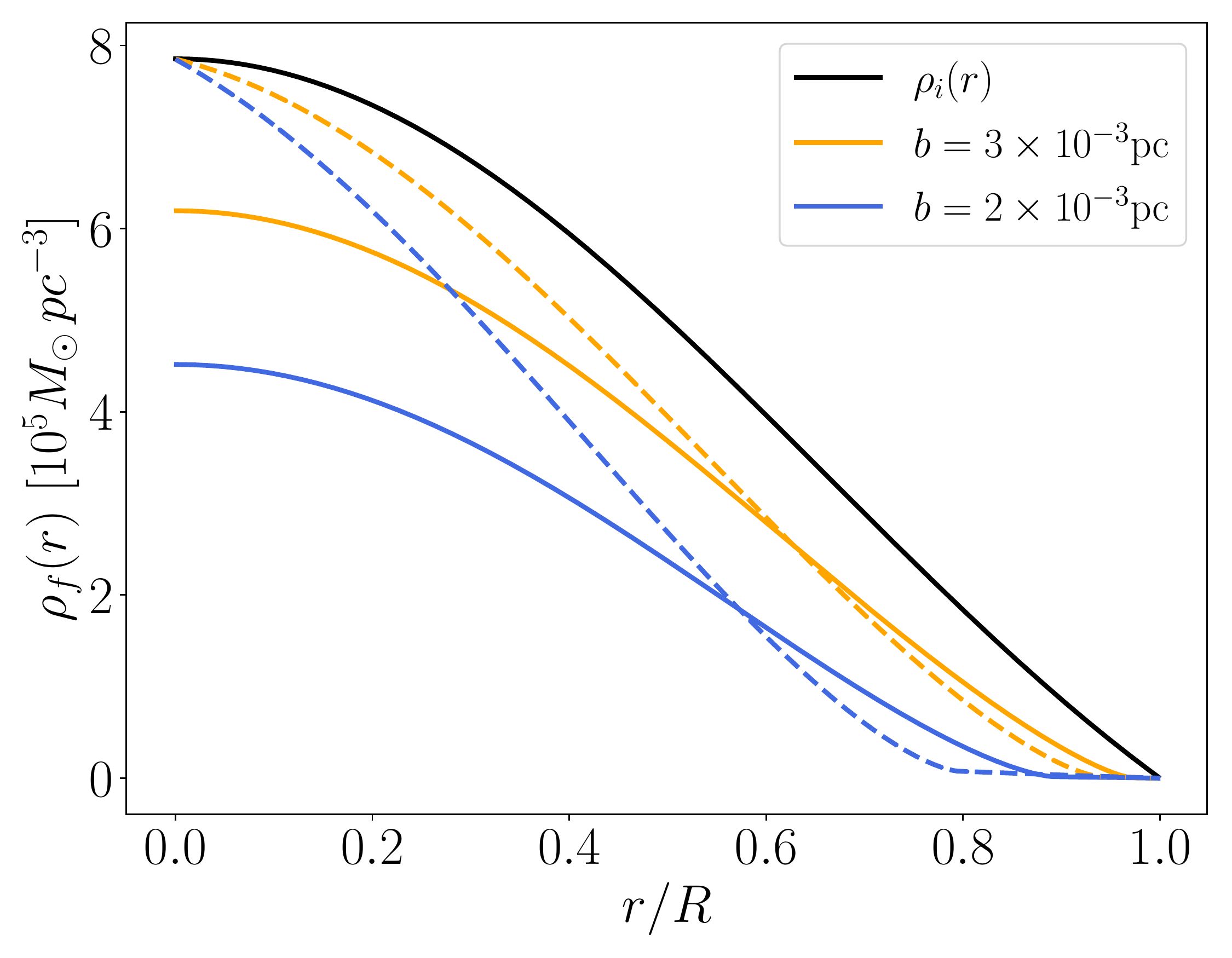}
  \includegraphics[scale=0.32]{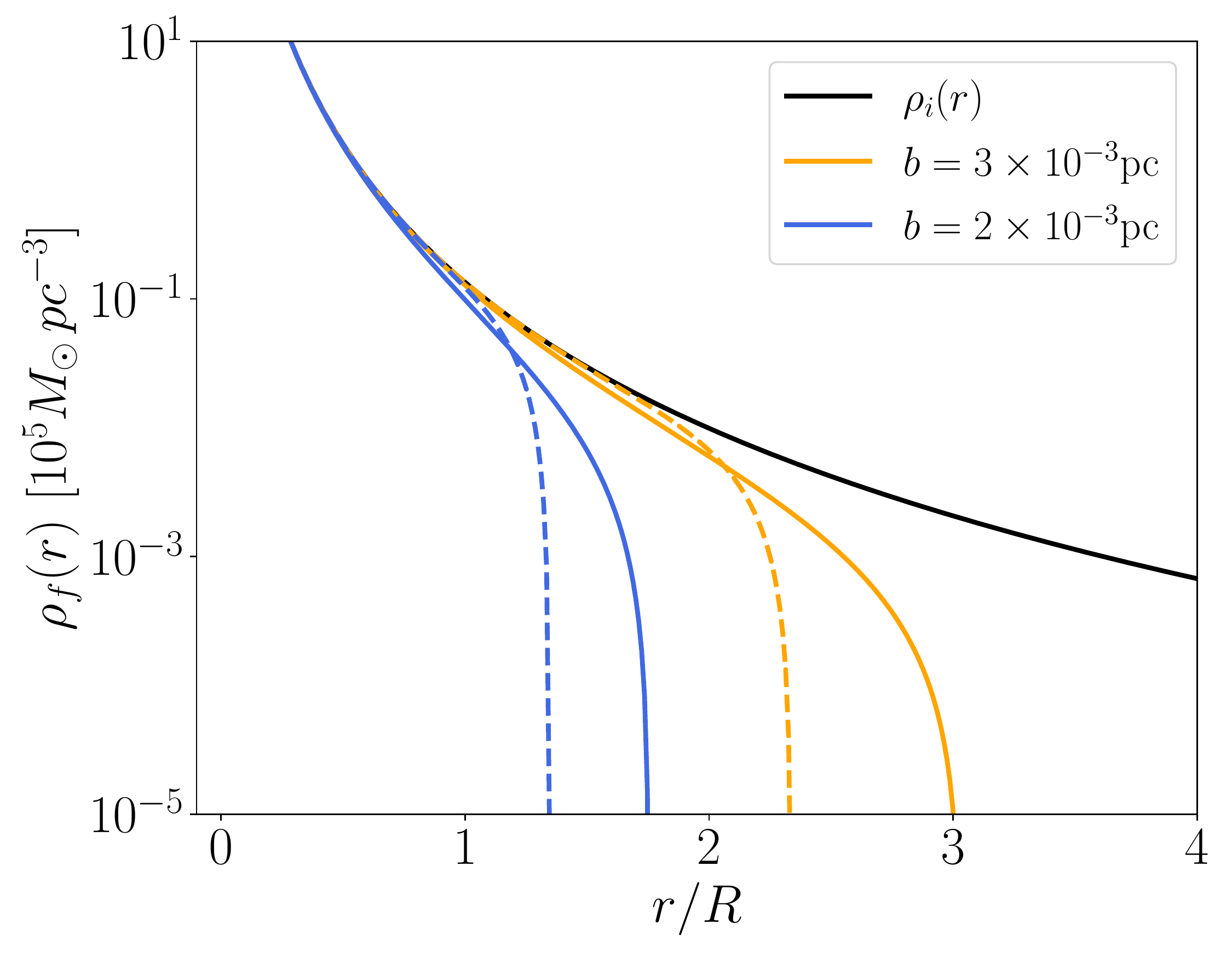}
 \caption{Density profile $\rho_f = \rho_i - \Delta \rho$ just after the interaction for miniclusters with Lane-Emden (left) and Hernquist (right) profiles, where $\rho_i$ is the density profile before the encounter. In both cases the initial masses and radii are set to $M=10^{-12}\,M_\odot$ and $R=10^{-6}$~pc. The final density profile $\rho_f(r)$ is shown for two different impact parameters: $b=2\times10^{-3}$~pc (blue), $b=3\times10^{-3}$~pc (orange). The dashed lines represent the same calculations done in the particle picture. The black curves show the profile before the perturbation. The parameters of the star are set to $M_* = 1\,M_\odot$ and $v = 10^{-4}$.}
\label{fig:Drho}
\end{figure}

The density profile right after the interaction is shown in Fig.~\ref{fig:Drho} for different impact parameters. We find that there is a significant difference between our formalism (solid line) and what a calculation in a particle picture would return (dashed line). Indeed, since in the particle formalism, the energy received depends on the location, $\delta E(r)\propto r^2$, no particle will be removed close to the origin. In our wave formalism, the whole energy level is shifted with $\delta E(E,l)\propto \ev{r^2}_{E,l}$ such that the density profile will be affected close to the origin, as observed in Fig.~\ref{fig:Drho} (left).  For the Hernquist profile this effect is less pronounced because the density diverges for $r\to 0$. For both profiles, more mass is removed from large radii in the particle compared to the wave treatment. 

\bigskip

When the cluster loses some of its mass, its energy will also be modified. Firstly, the energy carried initially by the removed modes has to be subtracted from the system. This will lead to a change in the kinetic energy, $\Delta E_K$, and binding energy, $\Delta E_B$, of the system. The change in kinetic energy is given by\footnote{This expression can be obtained by starting from Eq.~\eqref{eq:K} for the total initial kinetic energy and using that within the WKB approximation we have
$\nabla^2 R_{nl}(r)Y_{lm}(\theta,\varphi) = -2m_a(E_n - m_a\phi(r)) R_{nl}(r)Y_{lm}(\theta,\varphi)$.}
\begin{equation} \label{eq:EK}
    \begin{split}
\Delta E_K = -16 \pi^2 m_a \int^\infty_0 dr  \int^{l_{\rm max}(r)}_0 dl\,l  
\int^0_{\Tilde{E}(l)} dE\,   (E-m_a\phi(r)) \frac{f(E)} {\sqrt{2m_a(E-V_l(r))}} \,,
    \end{split}
\end{equation}
while the variation of the binding energy is obtained as $\Delta E_B = W_f - W_i$, where the final and initial binding energies are calculated from Eq.~\eqref{eq:W} by using $\rho_{f,i}$, respectively, with $\rho_f = \rho_i - \Delta\rho$.
Hence, the removed modes modify the system energy as 
\begin{equation}
    \Delta E_{\rm lost}=\Delta E_K +\Delta E_B \,.
\end{equation}

Secondly, the energy of the system will be also modified by the energy injected by the star into the modes that stay bound to the system ($\Delta E_{\rm inj.}$), i.e., the modes with $E < -\delta E (E,l)$. Explicitly, if Eq.~\eqref{total injected energy} represent the total energy injected by the star, the part that would go to the states that stay bound to the clump is
\begin{equation}\label{eq:Einj}
    \Delta E_{\rm inj.} = 16 \pi^2 m_a \int^\infty_0 dr  \int^{l_{\rm max}(r)}_0 dl\,l  \int^{-\Tilde{E}(l)}_{E_{min(l)}} dE 
    \frac{f(E)\,\delta E(E,l)}{\sqrt{2m_a\left(E-V_l(r)\right)}}\,.
\end{equation}
In total, the final energy of the bound system is then given by\footnote{Note that $\Delta E$ defined in Eq.~\eqref{eq:DE} is the energy change of the \emph{bound system}, whereas $\Delta\mathcal{E}$ defined in section~\ref{sec:transfE} is the total energy transferred by the star. The latter is shared between states which become ejected with a fraction $f_{\rm ej}$ and states which remain bound with fraction $(1-f_{\rm ej}) = \Delta E_{\rm inj}/\Delta\mathcal{E}$. The energy carried away by the ejected part is $f_{\rm ej}\Delta\mathcal{E} - \Delta E_{\rm lost}$, such that 
$E_f + f_{\rm ej}\Delta\mathcal{E} - \Delta E_{\rm lost} = E_i + \Delta\mathcal{E}$, which is equivalent to Eq.~\eqref{eq:DE}.
See also Ref.~\cite{Kavanagh:2020gcy}.} 
\begin{equation}\label{eq:DE}
    E_f=E_i +\Delta E_{\rm lost} + \Delta E_{\rm inj.}  = E_i + \Delta E\,.
\end{equation}
Let us add the following comment at this point. The way the energy is modified can be interpreted as a competition between cooling and heating the system. Indeed, when ejecting some modes, the kinetic energy decreases by $\Delta E_K$ and the system is cooled down. On the other hand the system will be heated up by the variation of the binding energy $\Delta E_B$ and the injected energy $\Delta E_{\rm inj.}$. Hence, we may expect that whether the cooling or the heating process is dominant, the system will contract or expand itself toward a more dense or more diffuse object, respectively.

In order to describe the system after the encounter we follow Ref.~\cite{Kavanagh:2020gcy} and assume that the cluster will relax to a virialized state, while keeping its total energy $E_{\rm tot} = E_f$ constant.\footnote{More precisely, the system will undergo a series of contractions and expansions that take between 10 and 20 times the crossing time of the halo before going back to an equilibrium configuration \cite{Gnedin:1999}. The exact relaxation process in the wave formalism is beyond the scope of this work and is left for further investigations.}
This assumption is justified by noting that the relaxation time scale of the cluster is short compared the typical time between star encounters \cite{Kavanagh:2020gcy}. Furthermore, we assume that the clump will adopt the same density profile as before the encounter, i.e., either the Lane-Emden or the Hernquist profile. Under these assumptions the new radius after the encounter is determined by the relation in virial equilibrium, $E_{\rm tot} = W/2$ and the expressions for the binding energy given in Eq.~\eqref{eq:W_LE-H} for the two example profiles. Hence, the radial parameters for the Lane-Emden and Hernquist cases are given, respectively, by
\begin{equation}
\begin{split}
    R_f &= -\frac{3}{8}\frac{G\left(M-\Delta M \right)^2}{E_f} = \kappa R_i \,,\\
    r_{H,f} &= -\frac{1}{12}\frac{G\left(M-\Delta M \right)^2}{E_f} = \kappa r_{H,i} \,,
\end{split} 
\qquad\qquad \kappa \approx 1 - 2\frac{\Delta M}{M} + \frac{\Delta E}{|E_i|} \,,
\end{equation}
where $\Delta E = \Delta E_{\rm lost} + \Delta E_{\rm inj.}$ is defined in Eq.~\eqref{eq:DE}. 
Depending on the two competing terms in $\kappa$ the size of the cluster will either decrease or increase as a consequence of the interaction with the star.

\begin{figure}[t]
    \centering
    \includegraphics[scale=0.35]{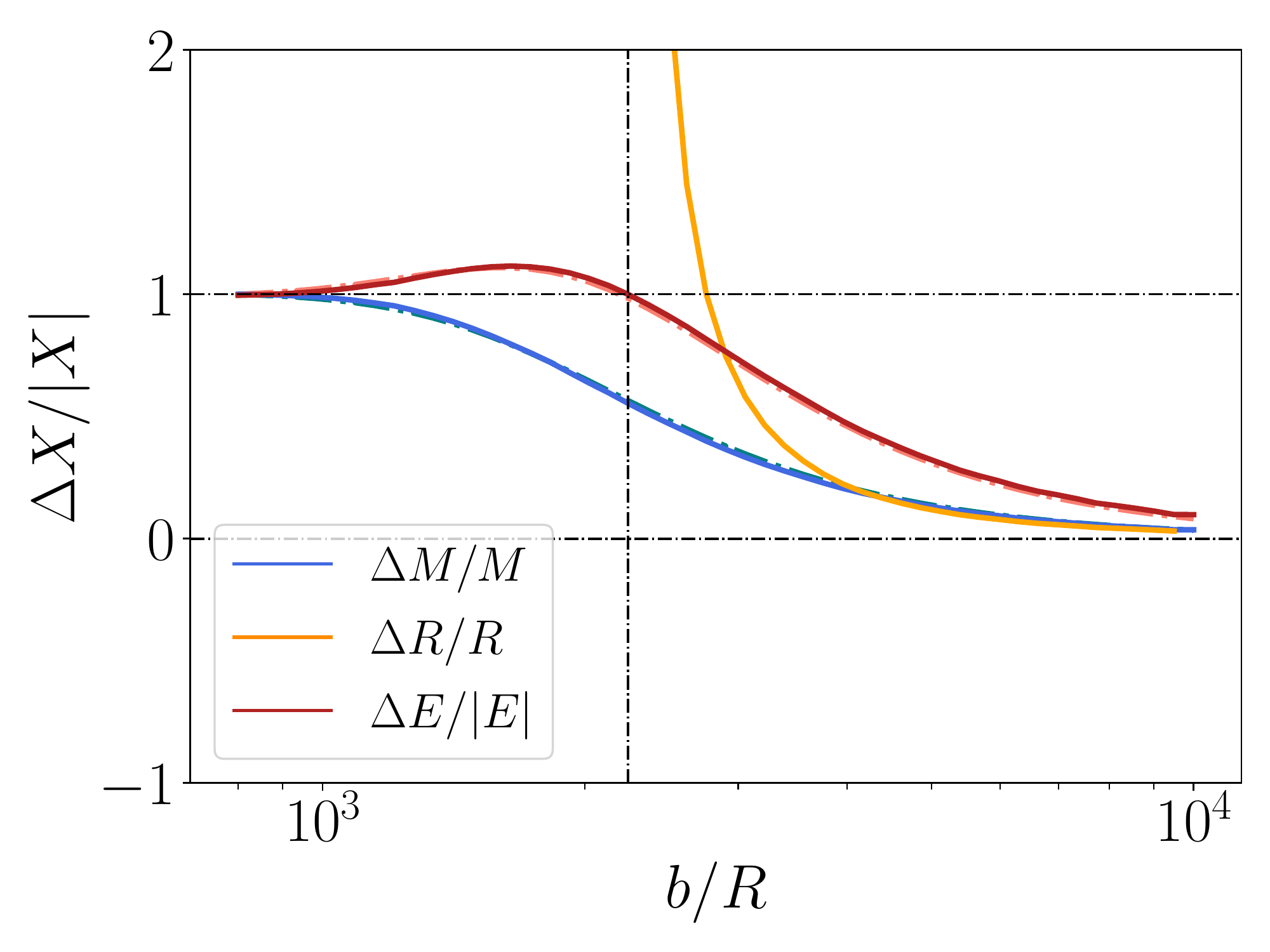}
   \includegraphics[scale=0.35]{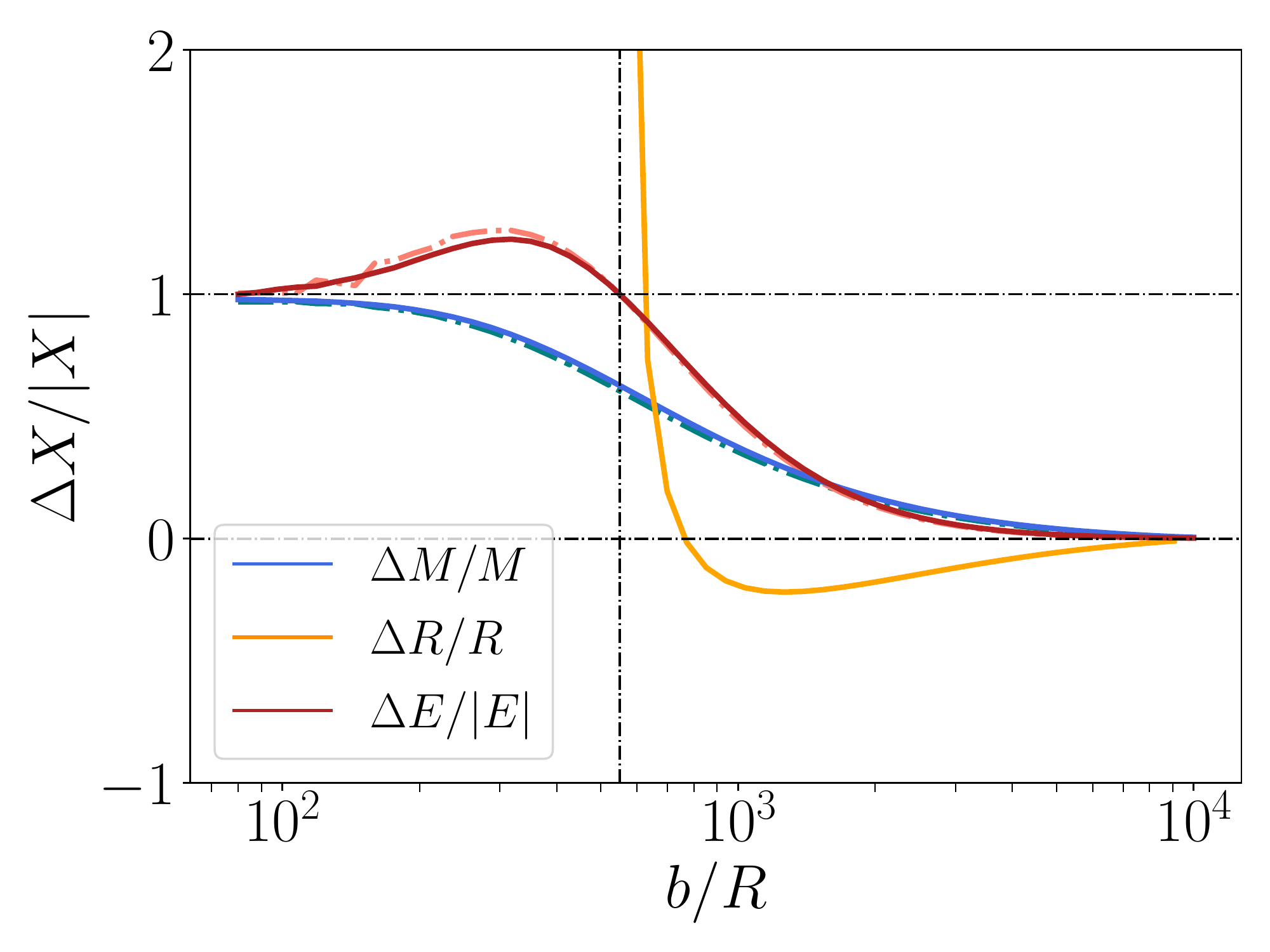}\\
   \includegraphics[scale=0.35]{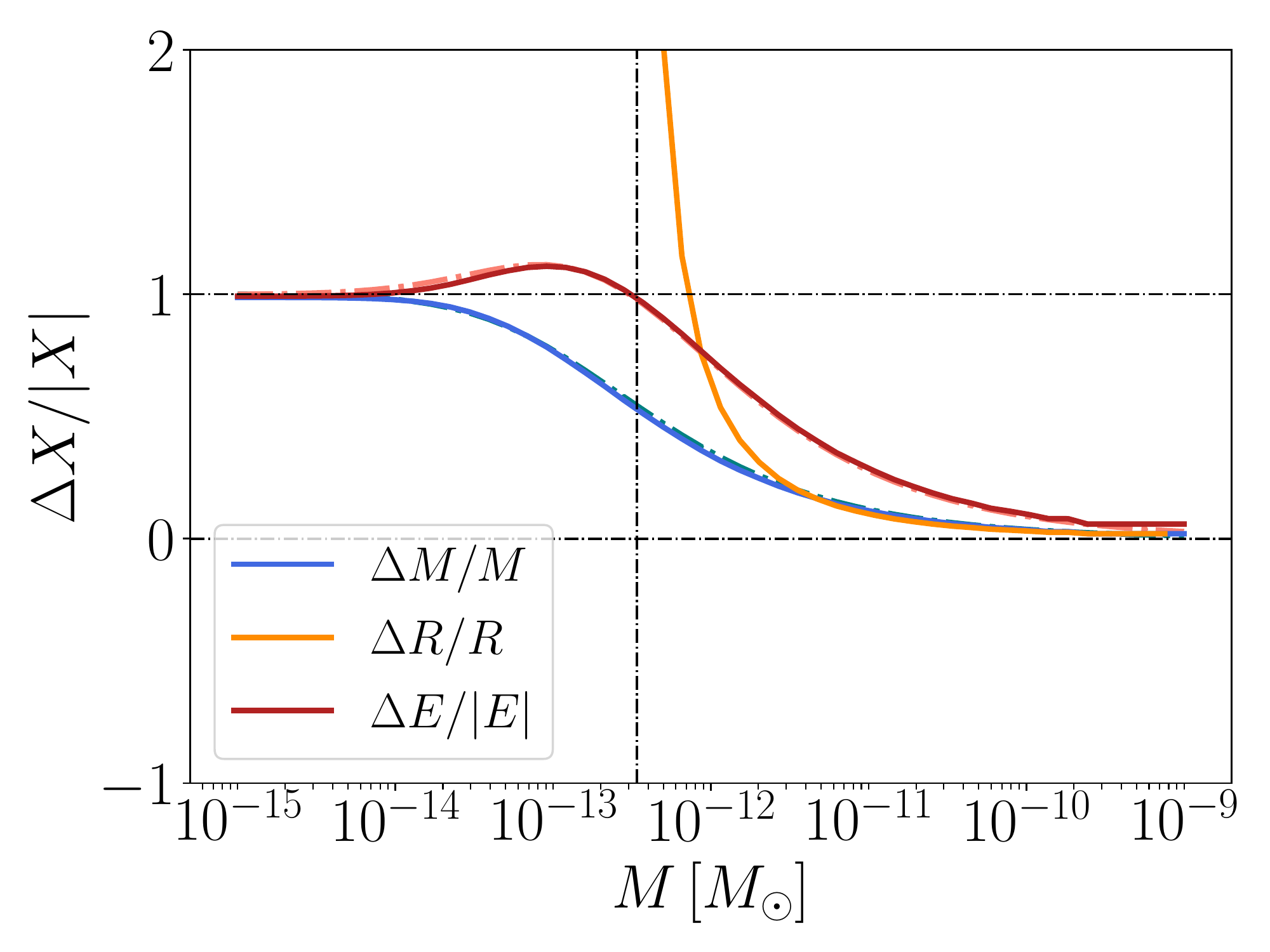}
   \includegraphics[scale=0.35]{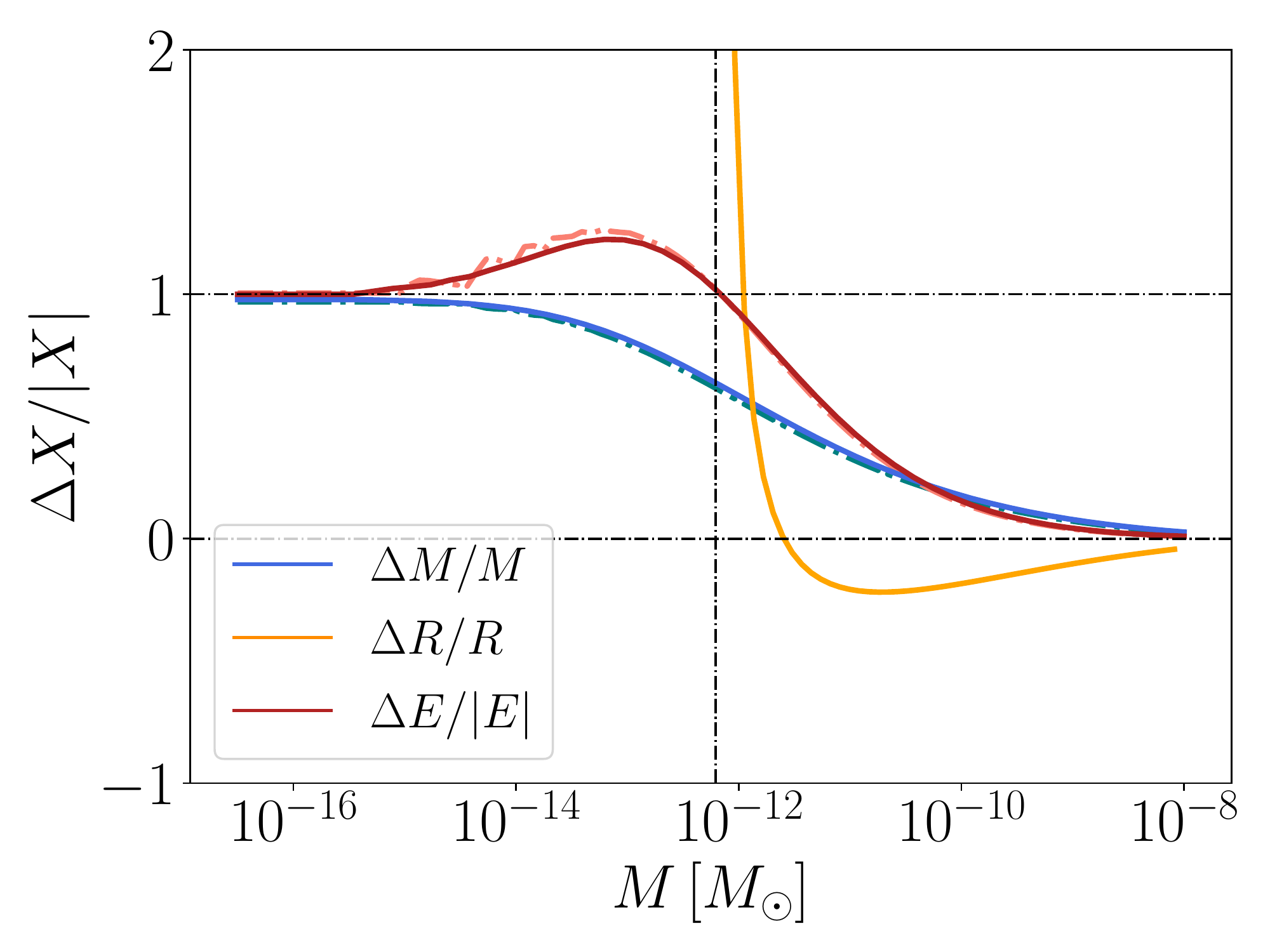}
   \includegraphics[scale=0.35]{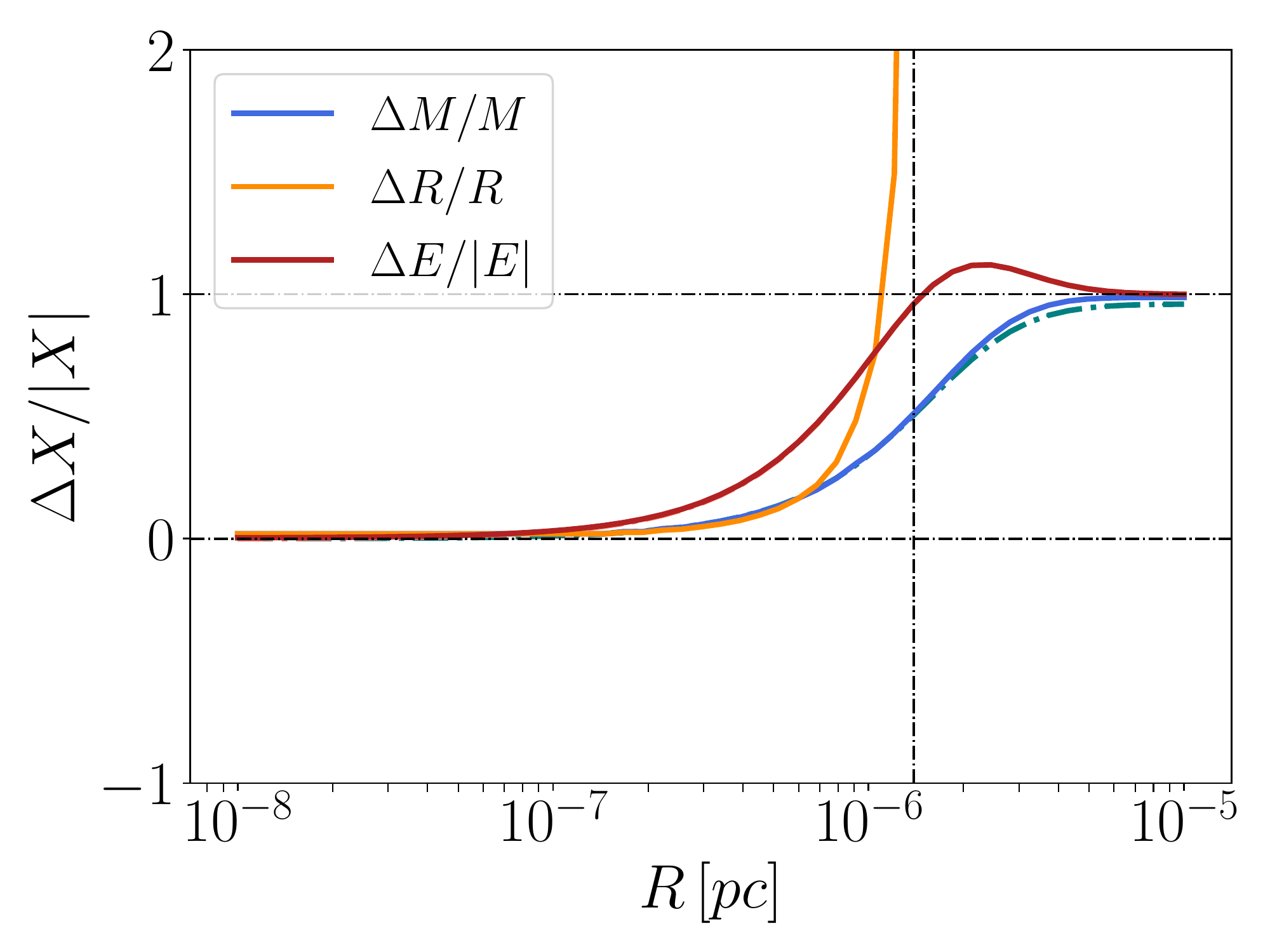}
   \includegraphics[scale=0.35]{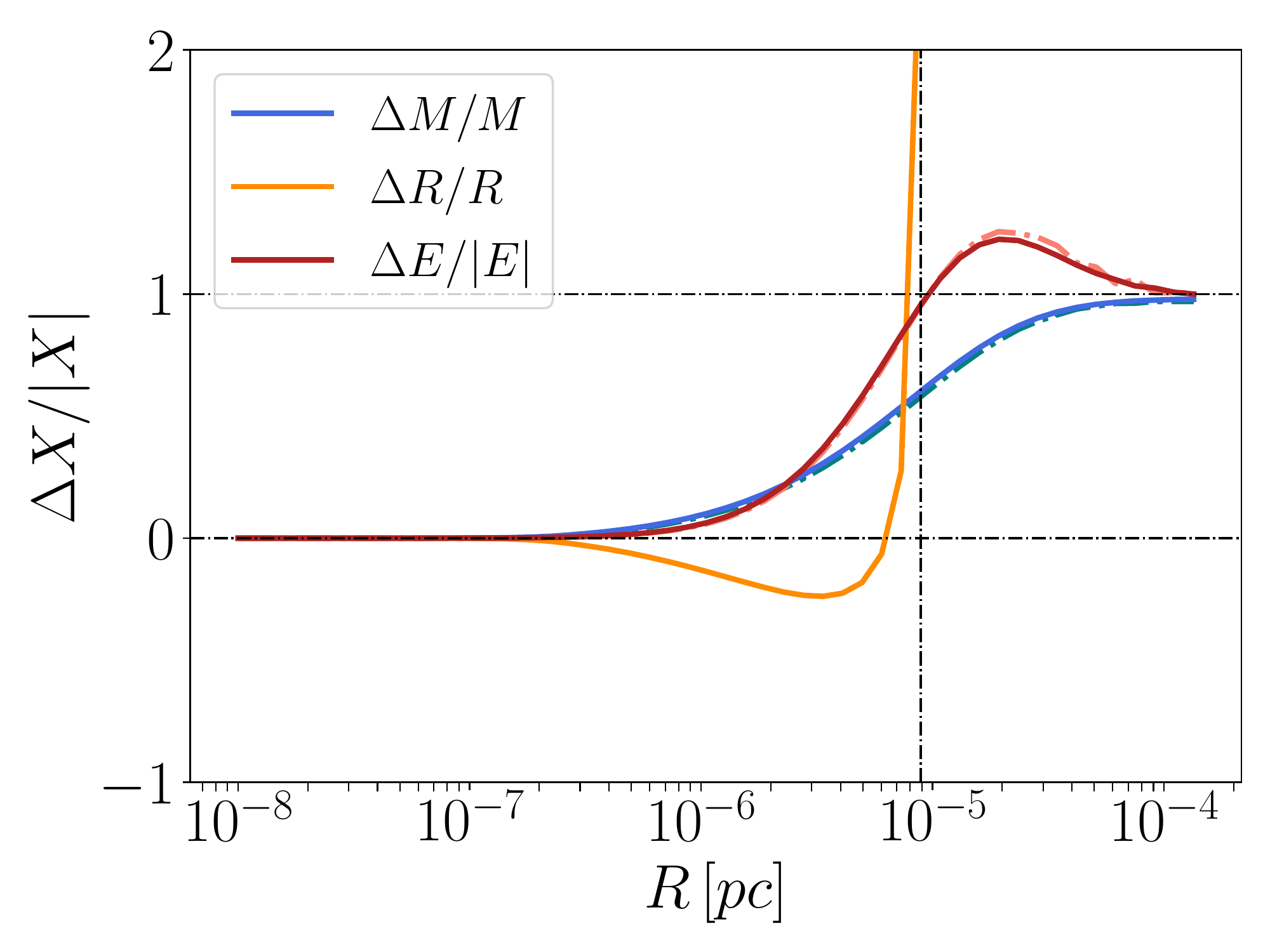}
   \caption{Relative change of the mass $\Delta M/M$ (blue), radius $\Delta R/R$ (orange) and total energy $\Delta E/|E_{\rm tot}|$ (red) for the LE (left panels) and H (right panels) profiles due to a star encounter. In the top panels the initial mass and radius are held constant and set to $M=10^{-12}\,M_\odot$, $R=10^{-6}$~pc while the impact parameter is modified. In the middle panels, the radius and the impact parameters are kept constant as $R=10^{-6}$~pc and $b=3 \, (0.5)\times 10^{-3}$~pc for LE (H). In the bottom panels the mass is kept fixed as $M=10^{-12}\,M_\odot$ and the impact parameter as $b=3\times 10^{-3}$~pc. Solid (dash-dotted) curves correspond to the wave (particle) description. The parameters of the star are set to $M_* = 1\,M_\odot$ and $v = 10^{-4}$. In the region to the left (right) of the vertical dashed line in the top, middle (bottom) panels $\Delta E/|E| > 1$, the perturbation is no longer small, and the clump is destroyed; we show these regions only for illustration purposes.}
    \label{fig:DX_X}
\end{figure}

In Fig.~\ref{fig:DX_X}, we apply the above formalism and study how the mass, radius, and total energy of the minicluster are affected by a star perturbation, as a function of the impact parameter (top), the minicluster mass (middle) and radius (bottom). For the LE miniclusters (left panels), after each interaction the system is heated up enough such that the radius increases whereas the mass decreases. This effect makes the cluster more vulnerable to further interactions leading to an eventual total destruction. This effect has also been observed in works using a particle formalism (for instance in Ref.~\cite{Kavanagh:2020gcy}, which adopts a different density profile, though). On the other hand, for miniclusters with an Hernquist profile (right panels), the distribution function assures a more diffuse density in the external layers of the clump. For small enough interactions, these layers are removed but the induced heating is not important enough to increase the radius afterward. As observed in Fig.~\ref{fig:DX_X}, for small enough perturbations, the radius decreases making the system more tightly bound and resistant to further interactions. However, if the perturbation is big enough, the more dense layers of the clump are removed and we end up in a similar situation as for the LE miniclusters, and the radius has to be increased to reach a new equilibrium configuration. The different behaviour of the radius change of gravitationally bound systems under tidal perturbations has been discussed e.g., in~\cite{Gieles_2016}. This reference finds a similar behaviour as we do: expansion (contraction) of weaker bound (stronger bound) systems.  

For both profiles, the energy of the system grows up to the point where the total energy becomes positive, i.e., for $\Delta E/|E_{\rm tot}| = 1$. At that stage, the system after the encounter does no longer form a bound structure and hence, the cluster is destroyed. This is manifest in Fig.~\ref{fig:DX_X} by a divergent radius at that location. 
In the middle/bottom panels of Fig.~\ref{fig:DX_X}, we fix the impact parameter and let the mass/radius change. As expected, as the mass increases or the radius decreases, the system is more tightly bound and it becomes more and more difficult to extract mass from it. This effect will be important for the survival of the miniclusters as a function of their mass/radius as discussed below. 
Indeed, we observe that the survival for a given impact parameter depends only on the average density of the cluster\footnote{We remind the reader that we define $R$ for the Hernquist profile as the radius which contains 80\% of the mass, whereas for the Lane-Emden case it contains 100\%.}, $\bar\rho = 3M/(4\pi R^{3})$ \cite{Kavanagh:2020gcy}. Numerically we find the critical density, below which the cluster gets destroyed of
\begin{equation}\label{eq:rho_crit}
\bar\rho_{\rm crit} \approx \left(\frac{M_*}{1\, M_\odot}\right)^2
\left(\frac{10^{-3} \, {\rm pc}}{b}\right)^4
\left(\frac{10^{-4}}{v}\right)^2 \times \left\{
\begin{array}{l@{\qquad}l}
0.7 \times 10^{-11} \, M_\odot\,(10^{-6}\, {\rm pc})^{-3} & \text{(Lane-Emden)} \\
2.1 \times 10^{-14} \, M_\odot\,(10^{-6}\, {\rm pc})^{-3} & \text{(Hernquist)}
\end{array}
\right. \,.
\end{equation}
The parameter dependence can be understood from Eq.~\eqref{injected energy}. We observe that Lane-Emden clumps are destroyed for densities about 3 orders of magnitude larger than the Hernquist clumps.


Finally we compare in Fig.~\ref{fig:DX_X} the changes in mass and total energy derived from the wave formalism (solid curves) to a corresponding calculation in a particle picture (dash-dotted curves), following the method applied in Ref.~\cite{Kavanagh:2020gcy}. We observe a remarkable agreement between the two methods, as the two sets of curves are hardly distinguishable in the figures. Since the final radius is a function of mass and energy after the encounter, also the radius will behave very 
similar in the particle formalism. 
This agreement seems to be a numerical coincidence, as we have observed in Fig.~\ref{fig:Drho} a notable difference between wave and particle description in the density profile right after the interaction. It remains an interesting question, whether this difference may become more important in a more realistic treatment of the virialization process. 

\begin{figure}[t]
\centering
  \includegraphics[scale=0.36]{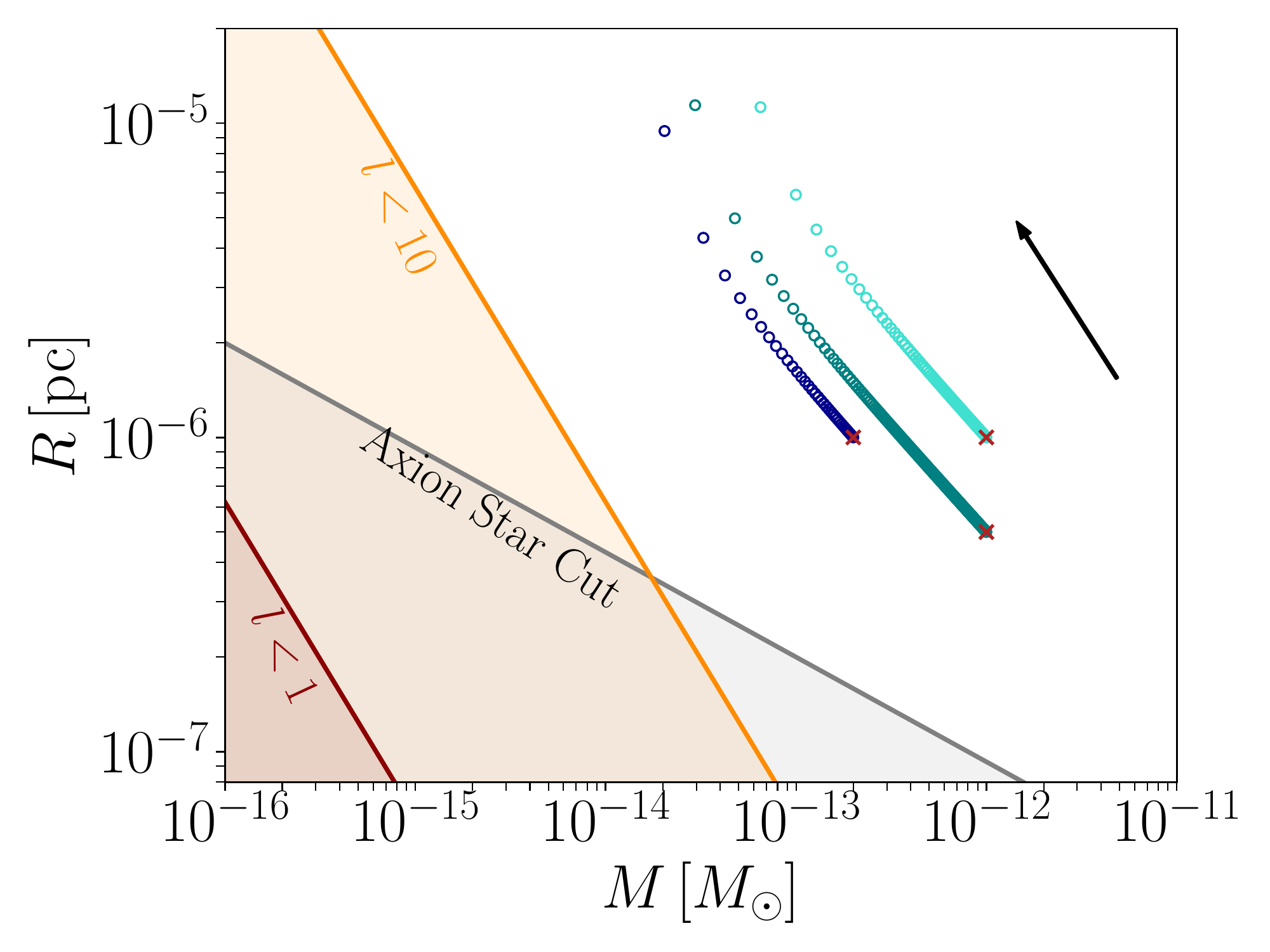}
  \includegraphics[scale=0.36]{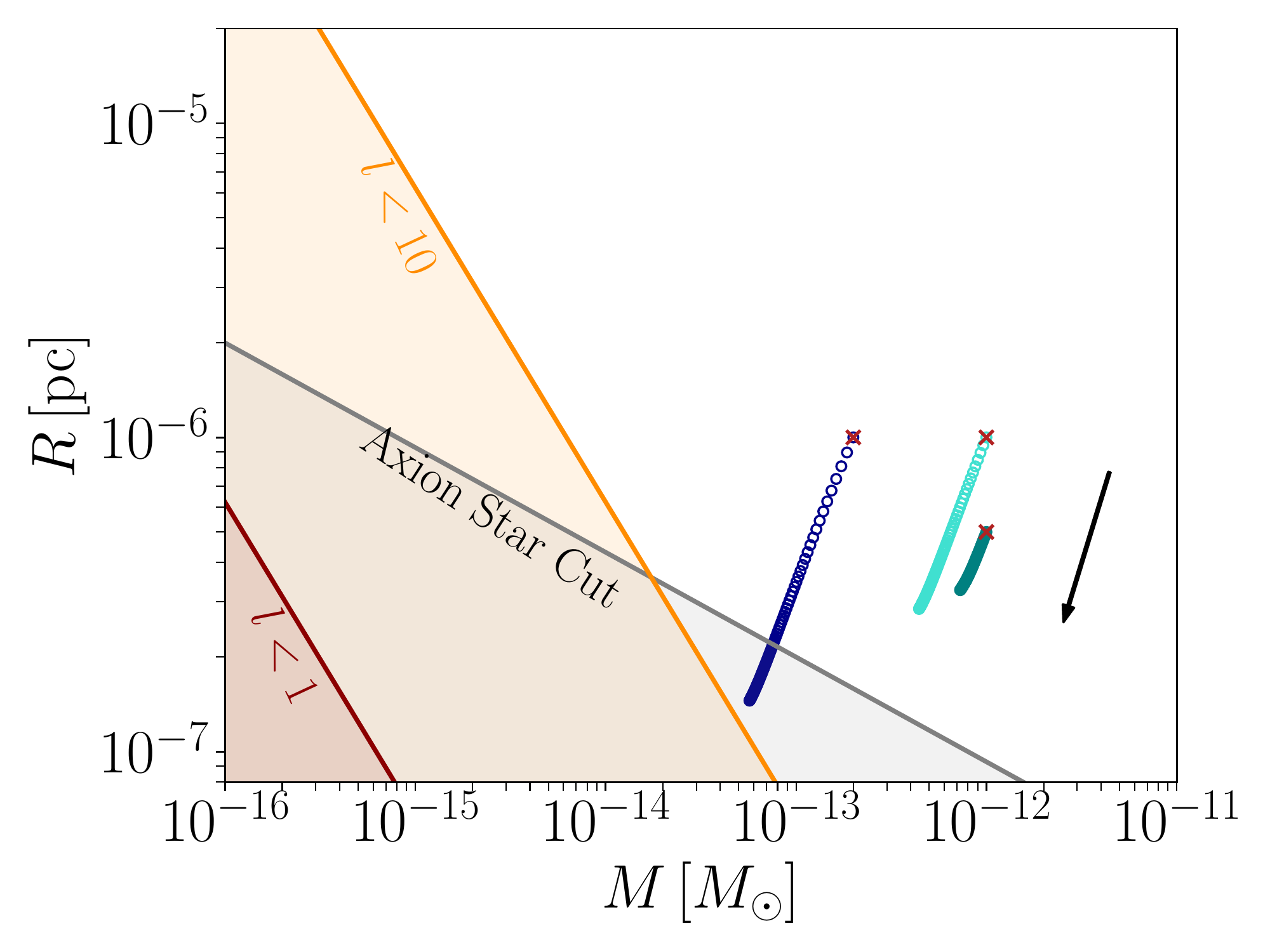}
   \caption{Evolution of the LE (left) and H (right) clumps under the same repeated perturbation: $b=2\, (0.5)\times 10^{-2}$~pc for the LE (H) clump, $v=10^{-4}$ and $M_* = 1\, M_\odot$. Different initial conditions have been used (red crosses) in order to underline that higher mass and smaller radius correspond to more resistant clusters. The total number of perturbations is fixed to $N=100$ for the H clumps whereas for the LE ones we let the perturbations continue until the cluster is destroyed. The gray, orange and red shaded regions show respectively the axion star cut, the $l_{\rm max}<10$, and the $l_{\rm max}<1$ regions.}
\label{fig:Evolution Minicluster}
\end{figure}

In Fig.~\ref{fig:Evolution Minicluster} we follow the evolution in the mass/radius plane of three  miniclusters with different initial conditions under repeated perturbations, keeping impact parameter, star mass and velocity constant. For the Lane-Emden clumps (left panel), each perturbation increases the size of the cluster while decreasing its mass, i.e., the system becomes less tightly bound. The impact of the perturbation increases and at some point the cluster is destroyed.  
In contrast, for the Hernquist clusters (right panel), the radius decreases and the system is more strongly bound after the perturbation and the effect of the next perturbation decreases. Hence, it becomes more difficult to reach the destruction point for clumps in that profile. In the figure
we show a finite number of $100$ interactions for the Hernquist case.

\subsection{Parameter space region of validity of our assumptions}
\label{sec:validity}

In Fig.~\ref{fig:Evolution Minicluster} we illustrate also regions in the mass/radius plane, where we expect that our description of miniclusters becomes unphysical. There are two main effects to be considered, namely the presence of an axion star inside the minicluster and the break-down of the WKB approximation.

As observed in numerical simulations,  \cite{Schive:2014hza,Levkov:2018kau,Eggemeier:2019jsu,Chen:2020cef}, typically a soliton (or axion star) \cite{Seidel:1993zk,Kolb:1993zz} forms in the center of miniclusters. 
The mass of the axion star ($M_{\text{AS}}$) is related to the minicluster mass $M$ via the so-called core--halo relation \cite{Schive:2014hza,Eggemeier:2019jsu}:
\begin{align}\label{eq:M_AS}
    M_{\rm AS} \simeq 1.3\times 10^{-17} \, M_\odot 
    \left(\frac{10^{-5} \, {\rm eV}}{m_a} \right) 
    \left(\frac{M}{10^{-13} \, M_\odot} \right)^{1/3} \,, 
\end{align}
which can be obtained by assuming that the virial velocities in the star and the halo are equal~\cite{Eggemeier:2019jsu}.
Using the definition of the radius of the axion star ($R_{\text{AS}}$) as a function of $M_{\rm AS}$ from Refs.~\cite{Schive:2014hza,Eggemeier:2019jsu}\footnote{We remark that the size $R_{\rm AS}(M_{\rm AS})$ used in
Refs.~\cite{Schive:2014hza,Eggemeier:2019jsu} is more than one order of magnitude smaller than the one obtained from the relation between radius and mass of an axion star derived from solving the Schr\"odinger--Poisson system, see e.g., \cite{Visinelli:2017ooc,Chavanis:2011zi}. The latter corresponds to the condition $l_{\rm max} \approx 1$, see below.}, we obtain \cite{Kavanagh:2020gcy}
\begin{equation}
\begin{split}
    R_{\text{AS}} &\simeq 3.4\times10^{-7}\,{\rm pc}     
    \left(\frac{10^{-5} \, {\rm eV}}{m_a} \right) 
    \left(\frac{M}{10^{-13}\,M_\odot}\right)^{-1/3} \,.
    \label{AS cut}
    \end{split}
\end{equation}
The grey shaded region below the line labeled ``Axion Star Cut'' in Fig.~\ref{fig:Evolution Minicluster} corresponds to the situation where the radius of the axion star is bigger than the radius of the minicluster. Under the assumption that the core--halo relation holds up to this point, we have to conclude that this region becomes unphysical, or at least the simple description in terms of superpositions of WKB solutions with random phases is no longer possible. Contrary, if we are sufficiently far away from that line, the presence of the axion star inside the cluster should be negligible for the behaviour of the cluster itself. 

Another requirement for the applicability of our formalism is the validity of the WKB approximation, Eq.~\eqref{eq:WKB}. As discussed in Sec.~\ref{sec:LE}, the WKB approximation is equivalent to the condition $l_{\rm max} \gg 1$, with $l_{\rm max}$ defined in Eq.~\eqref{eq:lmax}. The orange (dark red) shaded regions in Fig.~\ref{fig:Evolution Minicluster} indicate the parameter ranges where $l_{\rm max} < 10 \, (1)$. Hence, in these regions the deBroglie wavelength of the axion becomes comparable to the size of the cluster and the WKB solutions will no longer hold and we need to describe the system by close-to-ground state solutions of the Schr\"odinger-Poisson system. Indeed, numerically we find that the line $l_{\rm max} = 1$ is in good agreement with the axion star mass--radius relation found in Ref.~\cite{Visinelli:2017ooc,Chavanis:2011zi}. We note that in the region $l_{\rm max} \lesssim 10$ also a particle description of the system is no longer valid.

In the following discussion, we will restrict our analysis to the white region in Fig.~\ref{fig:Evolution Minicluster}, where the WKB wave description of the minicluster should be valid. We observe from the figure, that for the Lane-Emden profile, a given minicluster in the white region will always remain there under the action of star encounters, whereas the Hernquist cluster may be driven towards the ``forbidden'' region. Physically this means, that repeated perturbations may turn the initially diffuse minicluster into an axion star.

\bigskip

{\bf Comment on perturbativity.}
Let us point out that the formalism developed above is based on perturbation theory, assuming that the perturbation is ``small''. Requiring that the variation of the coefficients in Eqs.~(\ref{eq:D+},~\ref{eq:D-}) be smaller than $|C^{(0)}_{nlm}|^2$ gives us roughly the condition
\beq
\frac{A}{\alpha} \, R^2 = \frac{GM_*m_a R^2}{b^2v} < 1\,.
\eeq
For $M_*=M_\odot$ and $m_a=10^{-5}$~eV, the condition becomes
\beq
\frac{R}{b} < 5\times 10^{-3}\,v^{1/2}\,.
\eeq
We note that for some typical encounters in the galaxy this may be violated. However, we are mostly interested in the region where $\Delta E/|E|$ (as well as $\Delta M/M$) is small, c.f.~Fig.~\ref{fig:DX_X}. Although perturbativity may be violated for  some transitions, the overall effect of the perturbation on the clump is still small in this regime. When $\Delta E/|E|$ becomes of order one, we consider the clump as destroyed and do not further follow its evolution. Indeed, we never consider the region $\Delta E/|E| > 1$ (where the overall impact on the clump is large), which is shown in Fig.~\ref{fig:DX_X} only for illustration purpose.

\section{Survival in the galaxy}\label{sec:survival}

Several authors have studied the formation and initial properties of miniclusters in the post inflationary scenario, see, e.g.,  Refs.~\cite{Hogan:1988mp,Kolb:1995bu,Vaquero:2018tib,Buschmann:2019icd,Enander:2017ogx}. As a second stage in the history of these objects, collisions and merging after matter-radiation equality have to be taken into account to predict the halo mass function (HMF) at late times, 
see, e.g.,~\cite{Xiao:2021nkb, Ellis:2020gtq, Fairbairn:2017dmf, Fairbairn:2017sil, Eggemeier:2019khm, Ellis:2022grh}. Finally, once the dark matter halos become 
the sites of galaxy formation, tidal interactions with baryonic stars have to be introduced in the calculation to push the minicluster HMF up to the current days \cite{Tinyakov:2015cgg,Dokuchaev:2017psd,Kavanagh:2020gcy}. The purpose of this section is to address the last step using the formalism developed above, i.e., studying the evolution of a minicluster population inside the galaxy through tidal interactions with stars.

First, in section~\ref{sec:survival-MR}, we are going to calculate the minicluster survival probability at a given location in the Milky Way as a function of minicluster mass and radius. These results can then be applied for a given initial distribution of miniclusters. The minicluster distribution in mass and radius at late times is not well understood~\cite{Fairbairn:2017sil,Buschmann:2019icd,Sikivie:2006ni,Eggemeier:2019khm,Fairbairn:2017dmf,Ellis:2022grh}.
In section~\ref{sec:HMF}, we will adopt an initial HMF motivated by some of these studies, to give an illustrative example of how star encounters in the galaxy modify the HMF. We will discuss the differences between loosely and strongly bound axion miniclusters, using again the Lane-Emden and Hernquist profiles as respective examples.
In section~\ref{sec:assumptions}, we provide a discussion of the adopted assumptions and the limitations of our calculations.

\subsection{Survival probability as a function of the mass and the radius}\label{sec:survival-MR}

We want to calculate the survival probability of a particular minicluster with given mass $M$ and radius $R$.  As a first ingredient, we have to distribute the miniclusters in the galaxy in such a way that they correctly reproduce the NFW profile~\cite{Navarro:1996gj} for the dark matter halo of the Milky Way. This can be achieved by using specific distributions for the orbital parameters, semi-major axis $a$ and eccentricity $e$ \cite{Kavanagh:2020gcy}, see also \cite{Dokuchaev:2017psd}. 
We describe our procedure to obtain these distributions in Appendix~\ref{sec:NFWreconstruction}. The distribution of the eccentricity can be chosen independently of $r$ as 
$P(e) = 4e(1-e^2)$~\cite{vandenBosch:1998dc}. For given $r$ and $e$ we can then derive the distribution of the semi-major axis $P(a|r,e)$ as described in Appendix~\ref{sec:NFWreconstruction}. This allows us to calculate a distribution of $a$ and $e$ for the miniclusters which will reproduce the NFW profile of the halo.
For definiteness, we consider the location of the solar system in the Mily Way at $r \approx 8$~kpc.

The survival probability $P_{\text{surv}}(M,R,t,r)$ of miniclusters characterized by $M$ and $R$, after orbiting in the Milky Way for a time $t$, at a distance $r$ from the center of the galaxy, will be computed by creating for each pair $(M,R)$ an initial sample of $N^{(i)}_{\text{AMC}}(M,R,r)$ clusters with orbital parameters distributed as described above. Depending on the orbital parameters, each cluster in the sample will undergo a certain number of interactions. 
Consider a minicluster with a relative velocity $v$ compared to the perturbers and traveling for an infinitesimal duration $dt$. The number of stars with an impact parameter in the interval $[b,b+db]$ is given by \cite{Galactic}
\begin{equation}\label{eq:dN}
    dN = n_*(r) \times (v\, dt) \times (2\pi b \,db) \,,
\end{equation}
where $n_*(r)$ is the stellar density at the location of the cluster, which is assumed to be constant on the length scales of order of the relevant impact parameters. We use the angular averaged stellar density based on Refs.~\cite{Binney:1996sv,Bissantz:2001wx,Bensby:2004pc,SDSS:2005kst,Dehnen:1996fa,2011}, see Appendix \ref{sec:StellarPopulation} for details. The number of interactions per orbit is obtained by integrating Eq.~\eqref{eq:dN} and averaging over the distribution of relative velocities
\begin{equation}
    N_{\rm orb}(a,e) = \pi b_{\rm max}^2 \langle v \rangle 
    \int_0^{T_{\rm orb}(a)} dt \, n_*(r(t)) \,.
\end{equation}
The duration of one orbit as a function of the semi-major axis, $T_{\rm orb}(a)$, is given in Eq.~\eqref{OrbitPeriod}. We integrate up to a maximal impact parameter $b_{\rm max}$, which is chosen large enough such that encounters with $b>b_{\rm max}$ will have a negligible effect on the cluster.\footnote{Practically, we chose $b_{\text{\rm max}}(M,R)$ such that $\Delta M/M (b_{\text{\rm max}}) = 10^{-4}$. Pushing the maximum impact parameter to higher values would increase the number of interactions but would at the same time re-scale the probability distribution in Eq.~\eqref{eq:Pb}, such that the number of non-negligible interactions remains the same. We have checked that our results do not change when increasing $b_{\rm max}$.}
For the distribution of the relative velocities we take
\begin{equation}\label{eq:Pv}
  P(v) = \frac{4\pi v^2}{\left( 2\pi\sigma_\text{rel}^2\right)^{3/2}}\text{e}^{-\frac{v^2}{2\sigma_\text{rel}^2}} \,,
\end{equation}
with $\sigma_{\text{rel}} = \sqrt{2}\times10^{-3}$, which includes the velocity dispersion of the miniclusters and of the stars. The average number of encounters after a time $t$ is finally obtained as 
\begin{equation}
    N_{\rm tot}(t,a,e) = \frac{t}{T_{\rm orb}(a)} \, N_{\rm orb}(a,e) \,.
\end{equation}
We will simulate the evolution of the mass and radius of each cluster created in $N^{(i)}_{\text{AMC}}(M,R,r)$ by perturbing it $N_{\text{tot}}(t,a,e)$ times, drawing each time a velocity according to Eq.~\eqref{eq:Pv} and an impact parameter according to (compare Eq.~\eqref{eq:dN})

\begin{equation}
  P(b) = \frac{2b}{b_{\text{\rm max}}^2}\,.
    \label{eq:Pb}
\end{equation}

The surviving number of clumps at the end of the simulation $N^{(f)}_{\text{AMC}}(M,R,r,t)$ allows us to calculate the survival probability
\begin{equation}
    P_{\rm surv}(M,R,t,r)=\frac{N^{(f)}_{\text{AMC}}(M,R,r,t)}{N^{(i)}_{\text{AMC}}(M,R,r)}.
\label{eq:Survival}
\end{equation}
Importantly, miniclusters that cross either the WKB limit or the AS cut  line will also be removed from the sample since our formalism does not apply to them and we assume that they would form axion star-like objects.
We note that, after a minicluster is destroyed, the solitonic axion star core may survive.

\begin{figure}[t]
\centering
  \includegraphics[scale=0.37]{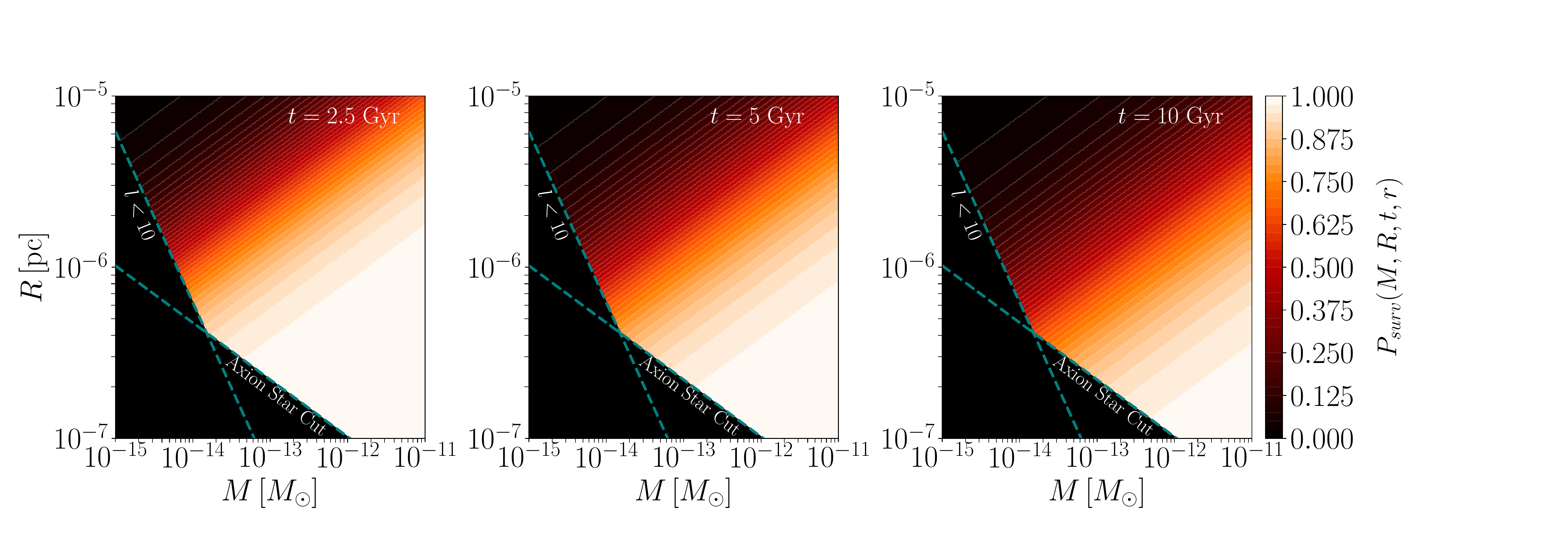}
   \includegraphics[scale=0.37]{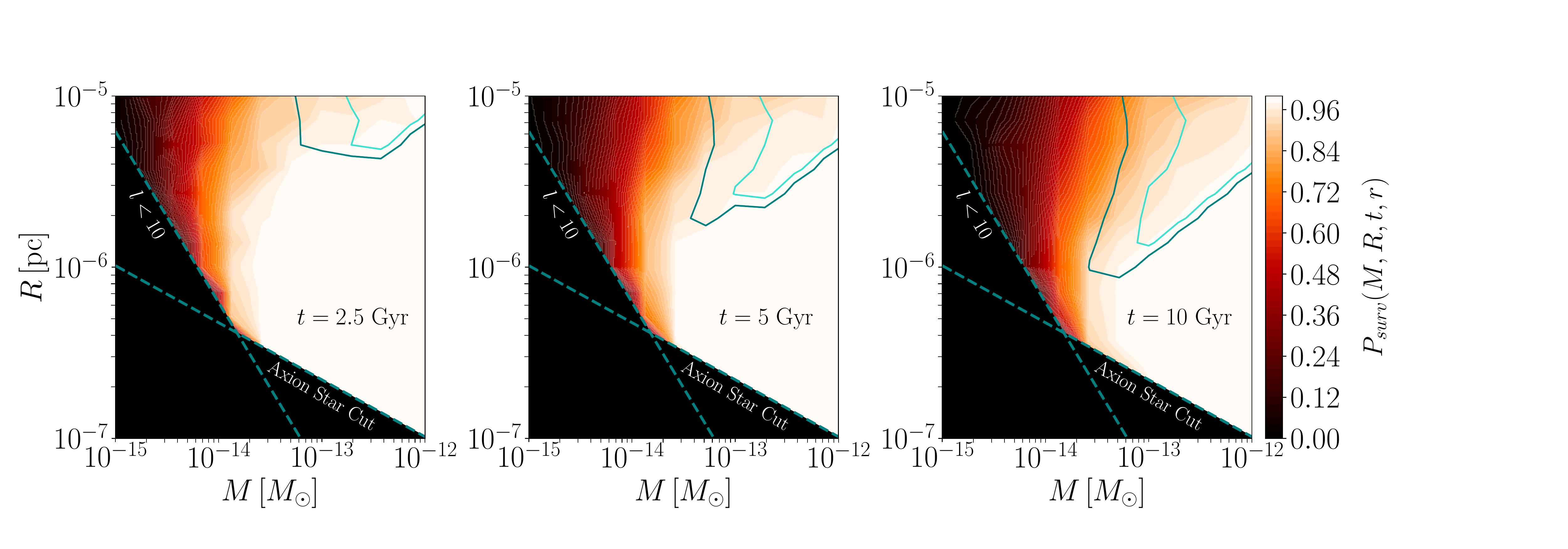}
 \caption{Time evolution of the survival probability $P_{\rm surv}(M,R,t,r)$ for the Lane-Emden (top) and Hernquist (bottom) miniclusters  at the sun location $r_\odot\approx 8$ kpc. The survival probability is shown for $t=2.5$~Gyr, 5~Gyr and 10~Gyr (from left to right). The axion star regimes are delimited by the blue dashed lines. For the Hernquist miniclusters we have included two contours showing the regions where 40\% (teal) and 80\% (turquoise) of the clusters that did not survive were actually destroyed instead of becoming an axion star. 
 }
\label{fig:Survival_MR}
\end{figure}

\begin{figure}[t]
\centering
 \includegraphics[scale=0.42]{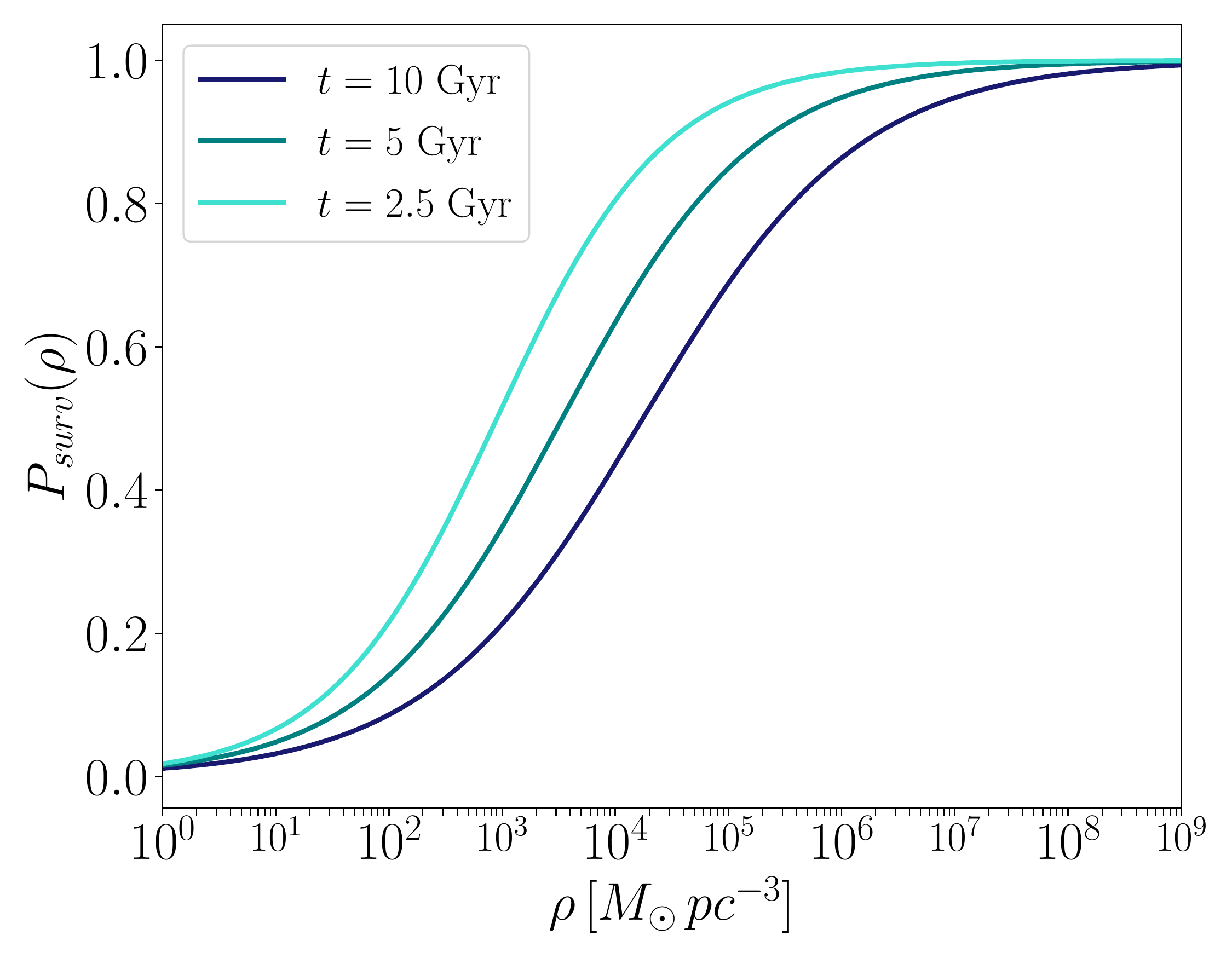}
 \caption{Local survival probability as a function of the initial mean density $\bar\rho$ for the Lane-Emden miniclusters. The three curves correspond to three different times.}
\label{fig:P_rho}
\end{figure}

Results are shown in Fig.~\ref{fig:Survival_MR}, where $ P_{\rm surv}(M,R,t,r)$ is shown at our location (the sun being located at $r_\odot\approx 8$~kpc from the center of the Milky Way) and for three different times. In the top panels, we assume the Lane-Emden profile. In agreement with the discussion in section~\ref{sec:perturbation}, we find that the survival probability of the miniclusters depends only on the mean density, with iso-survival-probability contours following $M\, R^{-3} = const$. In particular, the survival probability is less than $10\%$ for miniclusters with mean density lower than $\Bar{\rho} \approx (0.25,0.5,1.5)\times 10^{2} \, M_\odot \, \text{pc}^{-3}$ for $t=2.5,5,10$~Gyr, respectively. 
In Fig.~\ref{fig:P_rho} we show the survival probability as a function of the initial mean density for the same three different times. 
Since the radius always increases in the case of Lane-Emden clumps, no minicluster becomes an axion star during the interactions experienced in its lifetime, as they are driven further away from the axion star thresholds (see also Fig.~\ref{fig:Evolution Minicluster}). 


The survival probability for the Hernquist clumps is shown in the bottom panels of Fig.~\ref{fig:Survival_MR}, showing a rather different behaviour as the simple density-dependence found for LE. As discussed in the previous section, the radius of Hernquist clumps tends to decrease after each interaction, making the object more resistant to further perturbations. It is therefore very difficult for Hernquist clumps to be completely destroyed if the initial mean density is not small enough. However, since the radius is typically decreasing, the survival is affected close to the axion star and WKB cuts where the miniclusters could easily cross them and become an axion star. As we move to larger radii in Fig.~\ref{fig:Survival_MR} at constant $M$, the density becomes smaller, the system gets less bound and more affected by the perturbations. For this reason, it will also be possible for miniclusters to cross the axion star or WKB thresholds even if they are initially located further away from them, and we observe that larger regions to the right of the $l<10$ line become removed, as we move to larger radii. Even further away from the axion star thresholds, i.e., for large mass and large radius, we expect that also Hernquist clusters will get destroyed, c.f.~Fig.~\ref{fig:DX_X} or Eq.~\eqref{eq:rho_crit}. 
In that regime, the survival probability depends again only on the initial mean density, as for the Lane-Emden miniclusters.  This behavior is visible in the upper right parts of the Hernquist plots in Fig.~\ref{fig:Survival_MR}, where we indicate with two contours the fraction of the non-surviving clusters that get destroyed because $\Delta E/|E|$ becomes larger than 1 in an encounter.

\subsection{Impact of star encounters on the minicluster mass and radius distribution} \label{sec:HMF}

\begin{figure}[t]
\centering
 \includegraphics[scale=0.42]{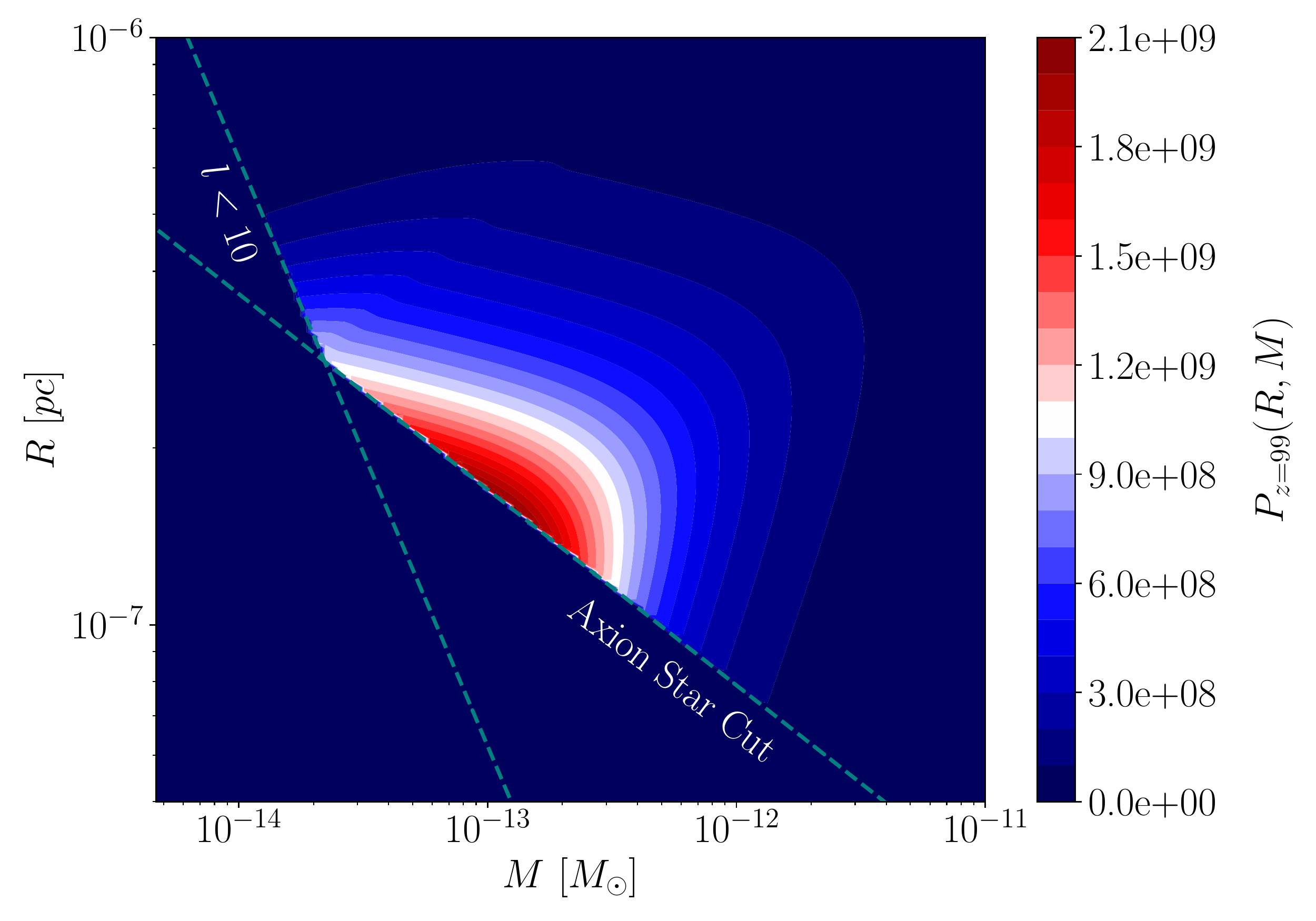}
 \caption{Mass and radius joint distribution $P_{z=99}(R,M)$ derived from Refs.~\cite{Eggemeier:2019khm, Buschmann:2019icd}, after having applied both the AS cut and the WKB condition.}
\label{fig:HMF}
\end{figure}

Equation \eqref{eq:Survival} can be used to calculate the absolute survival probability $P_{\rm surv}(t,r)$ of miniclusters at a given time and location for a specific model of the initial mass and radius distributions $P(R,M)$:
\begin{equation}
    P_{\rm surv}(t,r) = \int^{M_{\rm max}}_{M_{\rm min}} dM \int^{R_{\rm max}}_{R_{\rm min}} 
    dR\,P(R,M)\,P_{\rm surv}(M,R,t,r)\,,
    \label{Surv_General}
\end{equation}
where limits of the integrals correspond to the limits of the  mass and radius ranges according to the selected distribution.
In the following, we are going to use the mass and radius distributions obtained from numerical simulations~\cite{Fairbairn:2017sil,Buschmann:2019icd,Eggemeier:2019khm}. This is meant as an illustration to show the effect of star encounters on a given initial distribution. 
Letting the clusters merge and collide, simulations have shown that the mass distribution of these objects around $z=99$\footnote{The age of the Universe at $z=99$ is roughly 17~Myr.} should follow the expression~\cite{Fairbairn:2017sil,Eggemeier:2019khm}
\begin{equation}
    P_{z=99}(M)=\frac{\gamma}{M_{\text{\rm max}}^{\gamma}-M_{\text{\rm min}}^{\gamma}}M^{\gamma-1} \,,
    \label{Mass Distribution}
\end{equation}
with $\gamma=-0.7$, $M_{\text{\rm min}} =3.3\times 10^{-19}\,M_\odot$
and $M_{\text{\rm max}} =5.1\times 10^{-5}\,M_\odot$.
Secondly, a probability distribution $P(\delta)$ for the initial over-density $\delta$ that will lead to the cluster collapse has been found numerically in Ref.~\cite{Buschmann:2019icd}. If the initial over-density is related to the mean density of the clump by \cite{Kolb:1994fi}
\begin{equation}
    \Bar{\rho}(\delta)=140(1+\delta)\delta^3\rho_{\text{eq}} \,,
\end{equation}
where $\rho_{\text{eq}}$ is the average matter density at matter-radiation equality, for a given mass of the clump $M$, the radius is then given through
\begin{equation}
    R(M,\delta) = \left(\frac{3M}{4\pi\Bar{\rho}(\delta)}\right)^{1/3}.
\end{equation}
We assume that the distributions of $M$ and $\delta$ are independent.
Then, combining this with Eq.~\eqref{Mass Distribution}, we obtain a joint distribution in mass and radius, $P(M,R)$. The resulting distribution is also in rough agreement with the distributions derived in \cite{Enander:2017ogx} based on semi-analytic methods.

We note that the HMF according to Eq.~\eqref{Mass Distribution} peaks at the lower cut-off $M_{\rm min}$. It turns out that following this procedure to construct a mass/radius distribution, a large fraction of miniclusters would violate either the axion star (AS) cut or the WKB condition discussed in Sec.~\ref{sec:validity}. In order to restrict the analysis to objects for which our approach applies, we renormalise the distribution after applying both the AS and WKB cuts. We show the resulting joint probability distribution $P_{z=99}(M,R)$ in Fig.~\ref{fig:HMF}. Its integral over radius gives a new mass distribution $P_{z=99}(M)$, which now contains only miniclusters with an axion star having a smaller radius than the cluster itself and satisfying the WKB assumption.
\begin{table}[t!]
\centering
\begin{tabular}{ |l|c|c|c|  }
 \hline
 Density profile & $t= 2.5$ Gyr & $t= 5$ Gyr & $t= 10$ Gyr\\
 \hline
 Lane-Emden & $P_{\rm surv}= 94\%$  & $P_{\rm surv}= 90\%$ & $P_{\rm surv}= 82\%$ \\
 Hernquist & $P_{\rm surv}=99 \%$  & $P_{\rm surv}= 98\%$ & $P_{\rm surv}= 94\%$\\
 \hline
\end{tabular}
\caption{Survival Probability for the two different density profiles, Lane-Emden and Hernquist, at the Sun's location in the Galaxy as function of time. For the Lane-Emden profile the non-surviving clumps get distroyed by tidal interactions, whereas in the Hernquist case they become axion star-like objects.}
\label{Tab: Survival}
\end{table}

In Tab.~\ref{Tab: Survival}, we used Eq.~\eqref{Surv_General} to calculate the overall survival probability at different times according to the above distribution. We find that the survival probability for an initial population of Lane-Emden miniclusters is more affected by star encounters than the one for Hernquist clumps. After 10~Gyr about 18\% of LE clumps are destroyed, whereas only 6\% in the Hernquist case. Note that LE are driven away from the axion star configuration and eventually get destroyed, whereas 95\% of the removed Hernquist clumps in our simulation get removed because they cross the AS or WKB limits. By comparing the initial mass/radius distribution in Fig.~\ref{fig:HMF} and the lower panels in Fig.~\ref{fig:Survival_MR}, it becomes clear that the fraction of actually destroyed Hernquist clumps is small. As a result, we expect that a population of axion star-like objects is generated due to tidal interactions for Hernquist clumps.

\begin{figure}[t]
\centering
  \includegraphics[scale=0.36]{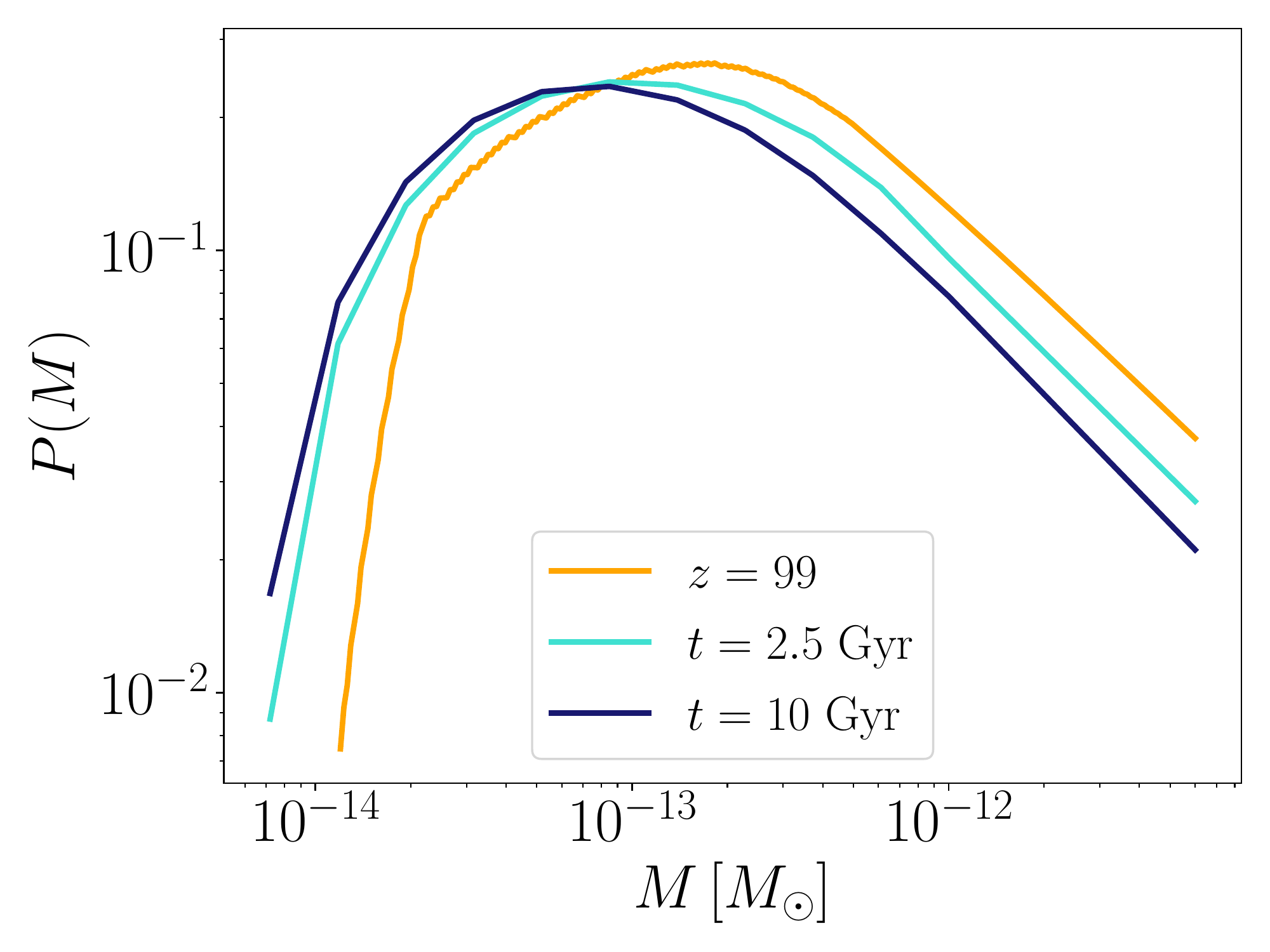} 
  \includegraphics[scale=0.36]{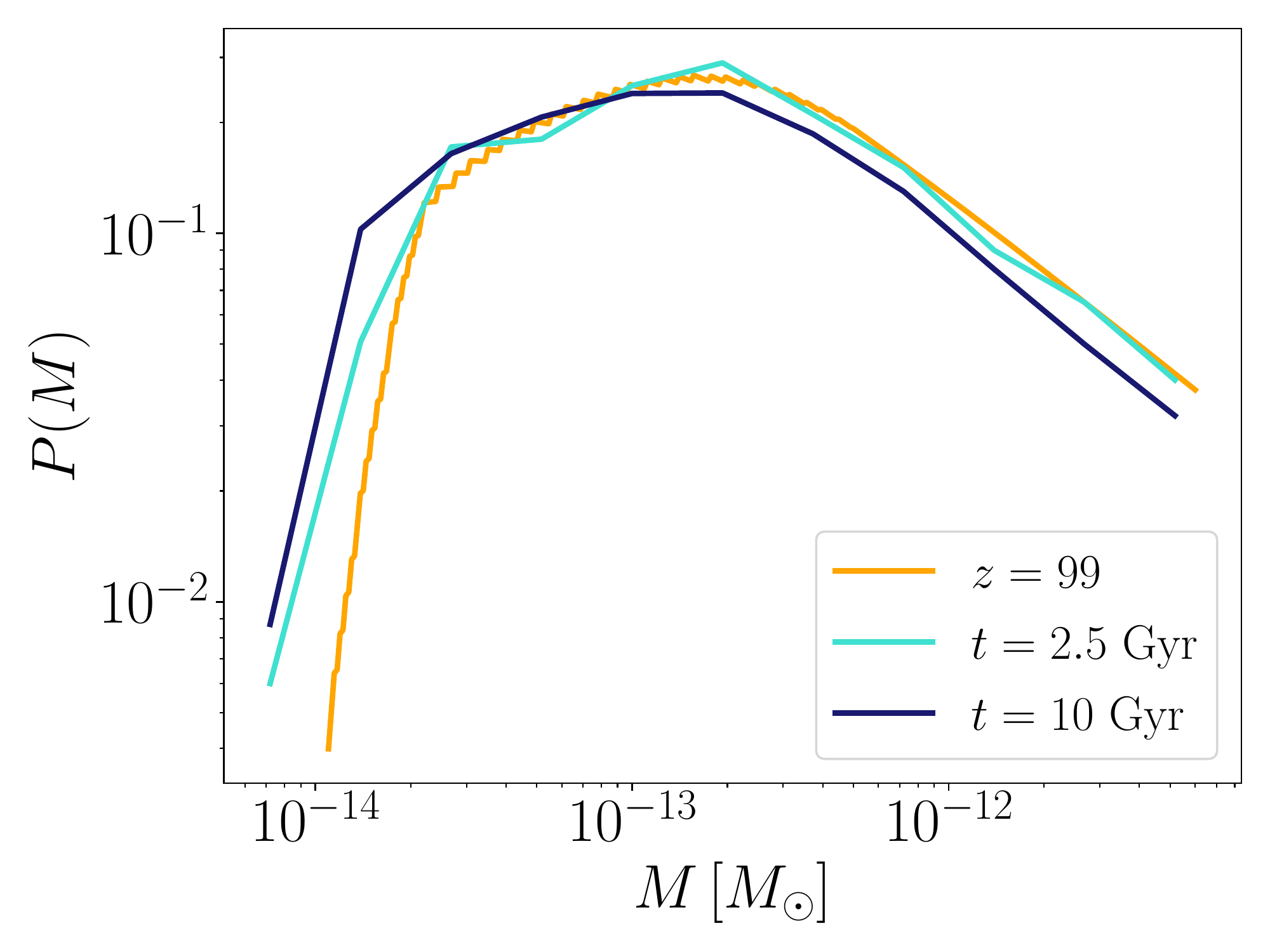}
 \caption{Evolution of an initial mass distribution (orange curve labeled ``$z=99$'') taking into account the tidal interactions with stars in the galaxy after 2.5 and 10~Gyr, at the Sun's location in the Milky Way. The normalization of the curves corresponds to the survival probability. 
 The left and right panels are respectively for the Lane-Emden and Hernquist miniclusters.}
\label{fig:Final HMF}
\end{figure}

Let us now consider the evolution of the mass distribution of the miniclusters. 
For a given minicluster with initial mass and radius $(M_i,R_i)$, we extract from the simulation the probability distribution $P_t(M|M_i,R_i)$ for the final mass $M$ after a time $t$. The normalization is such that the integral of $P_t$ over $M$ corresponds to the survival probability given in Eq.~\eqref{eq:Survival}. 
Starting from the initial mass/radius distribution described above, the mass probability distribution after a time $t$ is given by 
\begin{equation}
    P_t(M) = \int^{M_{\rm max}}_{M_{\rm min}}\int^{R_{\rm max}}_{R_{\rm min}}dM_i\,dR_i\,\, P_{z=99}(M_i,R_i)\,\,P_t(M|M_i,R_i) \,,
    \label{New HMF}
\end{equation}
which depends also on the location in the galaxy.
In Fig.~\ref{fig:Final HMF}, we show the time evolution of the mass distributions for both the Lane-Emden and Hernquist miniclusters.
In both cases we observe that the distribution is shifted to lower masses. As expected the effect is less important for the Hernquist clumps that are less sensitive to tidal interactions (see Tab.~\ref{Tab: Survival}).


\subsection{Assumptions and limitations}\label{sec:assumptions}

The previous calculations and results required some assumptions that we are going to summarize here. 
\begin{itemize}
    \item As introduced in Eq.~\eqref{eq:dN}, the number of interactions for a given minicluster depends on the density of stars  $n_*(r)$ along its orbit. We assumed here that the latter function is constant in time and given by its value today. A realistic approach would be to include the time dependence due to the birth and death of the stars in the Milky Way. As discussed in Ref.~\cite{Diehl:2006cf}, the rate at which stars are created is higher than the one at which they die, making the density of stars today an over-estimation of the real value at previous times. 
    
    \item For simplicity, we assumed throughout this paper that the whole population of perturbers has the same mass of $1\,  M_\odot$. We neglect the stellar mass distribution as well as the presence of other objects like white dwarfs or neutron stars. 
    But we note that variations in the stellar mass are largely degenerate with variations in $b$ and $v$.
    
    \item Similarly as for the stellar density, we assume the dark matter distribution in the Milky Way to be constant in time, up to time scales of 10~Gyr, ignoring the time evolution of the dark matter halo. In particular, considering an NFW profile, the evolution of the concentration as a function of redshift (related to the scale radius $r_s$) is derived in Ref.~\cite{Dutton:2014xda}, see also the Appendix of Ref.~\cite{Alvey:2021xmq} and references therein. This effect may have an impact on the results we have presented here since the distribution of miniclusters in the galaxy discussed in Appendix~\ref{sec:NFWreconstruction} would be now time dependent. Furthermore, the specific merger history of the Milky Way may lead to deviations from the NFW profile. A detailed investigation of these effects is beyond the scope of this work.
    
    \item We consider here a single source of disruption for the miniclusters -- through star interactions. A more realistic analysis would need to include  both the cluster mutual interactions and the stripping induced by the dark matter halo itself. 
    We note that the effect of the gravitational interaction scales as the perturbing object mass squared, see Eq.~\eqref{injected energy}. Since the mass of the clumps is of the order of $10^{-12} M_\odot$, we expect that the perturbation induced by a minicluster would be $10^{-24}$ times smaller than by a star with $1 M_\odot$. Noting further that the mass density of dark matter and stars in the sun's neighbourhood are roughly comparable, the number density of miniclusters is a factor $\sim M_{\odot}/M_{\rm MC}$ larger than the one of the stars, which compensates one power of that factor in the interaction, but still leaves a suppression of the overall effect of MC--MC interactions compared to the MC--star interactions of order $M_{\rm MC}/M_\odot \sim 10^{-12}$, and is therefore expected to be negligible \cite{Kavanagh:2020gcy}.
    
    However, the tidal stripping from the galactic halo might lead to significant mass modification of the miniclusters \cite{Jiang:2014nsa} even if the total disruption of the cluster is very unlikely \cite{Kavanagh:2020gcy}. We leave a detailed investigation of this effect for future work. The authors of Ref.~\cite{Dokuchaev:2017psd} estimate the collective effect of the gravitational potential of the galactic disk for miniclusters passing through the disk; they find that its effect is negligible compared to individual star encounters.
    
    \item An important assumption in our approach is, that after the encounter the minicluster re-virializes and  assumes {\it the same} density profile as before. This assumption ultimately needs to be checked by numerical simulations. It is an interesting question whether star encounters can actually lead also to changes in the shape of the density profile for the virialized object. 

   \item We ignore in our description the presence of the soliton (or axion star) inside the minicluster. As long as we are far away from the axion star cut line discussed above, this should be a good assumption, as the presence of the axion star should have a negligible effect on the behaviour of the minicluster, as both its radius as well as its mass are much smaller than the ones from the cluster, see eq.~\eqref{eq:M_AS}. However, as we approach the axion star cut line, the presence of the axion star may become important. This region is presently not well understood, and it is unclear whether the results of the numerical simulations of axion star formation inside the halo can be extrapolated to that point.
\end{itemize}

\section{Summary and discussion}\label{sec:conclusion}

We have presented a description of axion miniclusters in terms of a classical wave, based on WKB solutions of the Schr\"odinger equation. The cluster is built as a superposition of stationary solutions of the Schr\"odinger equation with random phases. The coefficients for the energy eigenstates are chosen according to a phase space density $f(E)$, which can be obtained from a given density profile $\rho(r)$ with the Eddington formula. In this way, a self-gravitating axion system can be obtained for any consistent pair of $f(E)$ and $\rho(r)$. As two specific examples, we discuss the Lane-Emden and the Hernquist profiles.

We have then used this formalism to study the effect of gravitational interactions of axion miniclusters with stars in the galaxy. The tidal perturbation can be formulated as a transition between energy levels by adopting standard quantum mechanics methods. Energy levels that are pushed to the continuum by the perturbation correspond to the mass stripped off by the star encounter. When the total energy of the system becomes positive the minicluster no longer corresponds to a bound system and is fully disrupted. We observe that whether the minicluster gets destroyed for given impact parameters depends only on the average density of the cluster. An important assumption in our analysis is that after the perturbation the minicluster re-virializes and settles to the same density profile as before the interaction. In this way, we can study the change in mass and radius due to the encounter. 
We compare our results in the wave formalism also to similar calculations using a particle description \cite{Kavanagh:2020gcy}. Although the density profile of the clump after the encounter is rather different in the two approaches, the global properties after re-virialization are very similar in the wave and particle pictures.

Finally, we perform a Monte Carlo simulation to study the survival of miniclusters in the Milky way, by assuming that the dark matter halo is initially made out of axion minicluster. We simulate a minicluster population in the halo assuming a semi-realistic distribution of orbital parameters for the clusters as well as the stellar density in the galaxy.

We find that the behaviour is rather different for the two example profiles. The radius of the Lane-Emden clump becomes larger after an encounter. This implies that the density decreases and the clump becomes more vulnerable to further perturbations. In this case, we find that the survival probability in the galaxy essentially depends on the average density of the cluster. Lane-Emden clusters with average density below few $10^2\, M_\odot {\rm pc}^{-3}$ are likely to be destroyed on times scales of few Gyr, c.f.~Fig.~\ref{fig:P_rho}. 
In contrast, for the Hernquist profile we find that typically the radius decreases after a stellar encounter, the system gets more strongly bound and becomes more resistant against further perturbations. In contrast to the Lane-Emden case (which has a finite density at $r=0$), for the Hernquist profile, the density goes as $1/r$ for small $r$. This leads to more tightly bound states in the central region.
For large $r$, the Hernquist profile drops as $1/r^4$, whereas Lane-Emden has a finite size. In typical star encounters with Hernquist clumps, the loosely bound states dominating at large radii are stripped off, leaving a tightly bound core, which after re-virialization settles to an overall smaller, more strongly bound system than initially. The new cluster becomes more stable against further perturbation. For typical encounters in the Milky Way, Hernquist clusters are hardly destroyed, but they are driven towards an axion-star like configuration, where our WKB ansatz breaks down, either because the deBroglie wavelength of the axion becomes comparable to the size of the cluster, or because the soliton configuration expected inside minicluster
\cite{Schive:2014hza,Levkov:2018kau,Eggemeier:2019jsu} becomes larger than the minicluster itself. Hence, we come to the interesting conclusion, that for profiles similar to Hernquist, we expect a population of axion stars to build up due to tidal interactions with stars. 

The different behaviour under tidal shocks of halos with different density profiles is known also from other gravitationally bound systems. Ref.~\cite{Gieles_2016} finds that Plummer and isochrone models, which have a $1/r^4$ behaviour at large $r$ (same as Hernquist), also contract under tidal perturbations and become more stable, whereas a Plummer profile with $1/r^5$ at large $r$ can become more dilute. Ref.~\cite{Kavanagh:2020gcy} considers a power law with $\rho \propto r^{-2.25}$ and an NFW profile with $\rho\propto r^{-3}$ at large $r$, and finds in both cases a behaviour similar to our Lane-Emden case. We note that the authors of Ref.~\cite{Kavanagh:2020gcy} truncate their profiles at a certain radius, which may have a similar effect as the finite radius of the Lane-Emden profile. Ref.~\cite{Dokuchaev:2017psd} assumes a rather shallow power law profile with $\rho(r) \propto r^{-1.8}$. Their survival probability is somewhat higher than our result for the Lane-Emden case, with their critical density being about one to two orders of magnitude lower than ours.
We also note that recent axion halo simulations \cite{Ellis:2022grh} cannot yet resolve the small $r$ region of the cluster and a power law with index $\approx 2.9$ provides a similarly good fit to the simulated halos as an NFW profile.

We leave it for future work to study in detail the behaviour under tidal perturbation as a function of the density profile properties or equivalently, the shape of the phase space density $f(E)$. Our results suggest that the density profile of axion miniclusters plays an important role in their fate under star encounters in the Milky Way. Hence, it is important to better understand the expected density profiles of axion minicluster in realistic cosmological scenarios.

\bigskip

 After the completion and submission of our work we became aware of Ref.~\cite{Yavetz:2021pbc}, where a similar method of constructing DM halos in the wave description has been discussed. Their method agrees with ours in the WKB limit.


\subsection*{Acknowledgement} 

We would like to thank Felix Kahlh\"ofer for comments and Javier Redondo for useful discussions. 
This project has received support from the European Union’s Horizon
2020 research and innovation programme under the Marie
Sklodowska-Curie grant agreement No 860881-HIDDeN.

\appendix

\section{Mapping between wave and particle description}
\label{app:mapping}

In this appendix we comment briefly on the mapping between the wave formalism developed in section~\ref{sec:axion-clump} and a description in terms of a phase space density $f_p(\vec{r},\vec{v})$ of point like particles with velocities $\vec{v}$ and coordinates $\vec{r}$. The subscript $p$ indicates that we are considering a system composed of particles. 

Consider the phase space density function corresponding to a stationary solution of the collisionless Boltzmann equation for a spherical system with isotropic velocity dispersion. As a consequence of Jean's theorem, such a distribution function can depend only the total energy of the particles: $f_p(\vec{r},\vec{v}) = f_p(E)$ with $E = m v^2/2 + m\phi(r)$~\cite{Galactic}. 
At fixed radius we have $m^2v^2 dv =\sqrt{2m(E - m\phi(r))}dE$, which is precisely the integration measure obtained in eq.~\eqref{eq:rho}. Therefore, we can identify the particle phase space density $f_p(E)$ with the density of states $f(E)$ for the energy eigenfunctions: 
$f_p(\vec{r},\vec{v}) = f_p(v,r) = f_p(E) = f(E)$.
Hence, eq.~\eqref{eq:rho} corresponds to the familiar form in terms of particle phase space density
\beqa
\rho(r) = 4\pi m^4\int_0^{v_{\rm max}(r)} dv\, v^2\, f_p(r, v)\enspace,
\eeqa
with $v_{\rm max}(r) = \sqrt{-2\phi(r)}$. 
With our chosen normalization, $f_p$ satisfies
\beq
m^3 \int d^3r\ d^3v\, f_p(\r,\v\,) = \frac{M}{m}\enspace,\label{phase_space_dens}
\eeq
where $M$ is the total clump mass.

\section{Virial theorem}
\label{app:virial}

The virial theorem in the context of the Schr\"odinger--Poisson system has been discussed e.g., in \cite{Chavanis:2011zi, Hui:2016ltb}. The starting point is the moment of inertia 
\begin{equation}\label{eq:I}
    I = \frac{m_a}{2} \int d^3r \, r^2 |\psi|^2 \,.
\end{equation}
Its time derivative can be obtained by using the
time dependent Schr\"odinger equation $i\partial_t \psi = H_0\psi$, with $H_0$ given in eq.~\eqref{eq:Schroedinger}. As shown in appendix~A of ref.~\cite{Hui:2016ltb} one can write the second time derivative as
\begin{equation}\label{eq:ddI}
    \Ddot{I} = -m_a\int d^3r |\psi|^2\r\cdot \vec{\nabla}\phi
    -\frac{1}{m_a}\int d^3r \psi^* \nabla^2\psi \,.
\end{equation}
Now consider the solution for $\psi$ constructed in section \ref{sec:formalism} and take the ensemble average of this equation. Using Poisson equation for a spherical system, the first term becomes the potential energy $W$ given in eq.~\eqref{eq:W}, whereas the second term becomes $2K$, with the kinetic energy $K$ defined by 
\begin{equation}\label{eq:K}
K = \left< -\frac{\nabla^2}{2m_a} \right>_{ens} \,.
\end{equation}
On the other hand, from eq.~\eqref{eq:I} we find
\begin{equation}
    \langle I\rangle_{ens} = \frac{1}{2} \int d^3r \, r^2 \rho(r) \,,
\end{equation}
which is time independent, c.f.~eq.~\eqref{eq:rho}. Therefore $\langle \Ddot{I} \rangle_{ens} = 0$ and the virial theorem $W+2K=0$ follows from eq.~\eqref{eq:ddI} for the ensemble average of our ansatz for the self-gravitating clump.

\section{Calculation of the perturbed coefficients}\label{app:coefficients}

\begin{figure}[t]
\centering
  \includegraphics[width=.8\linewidth]{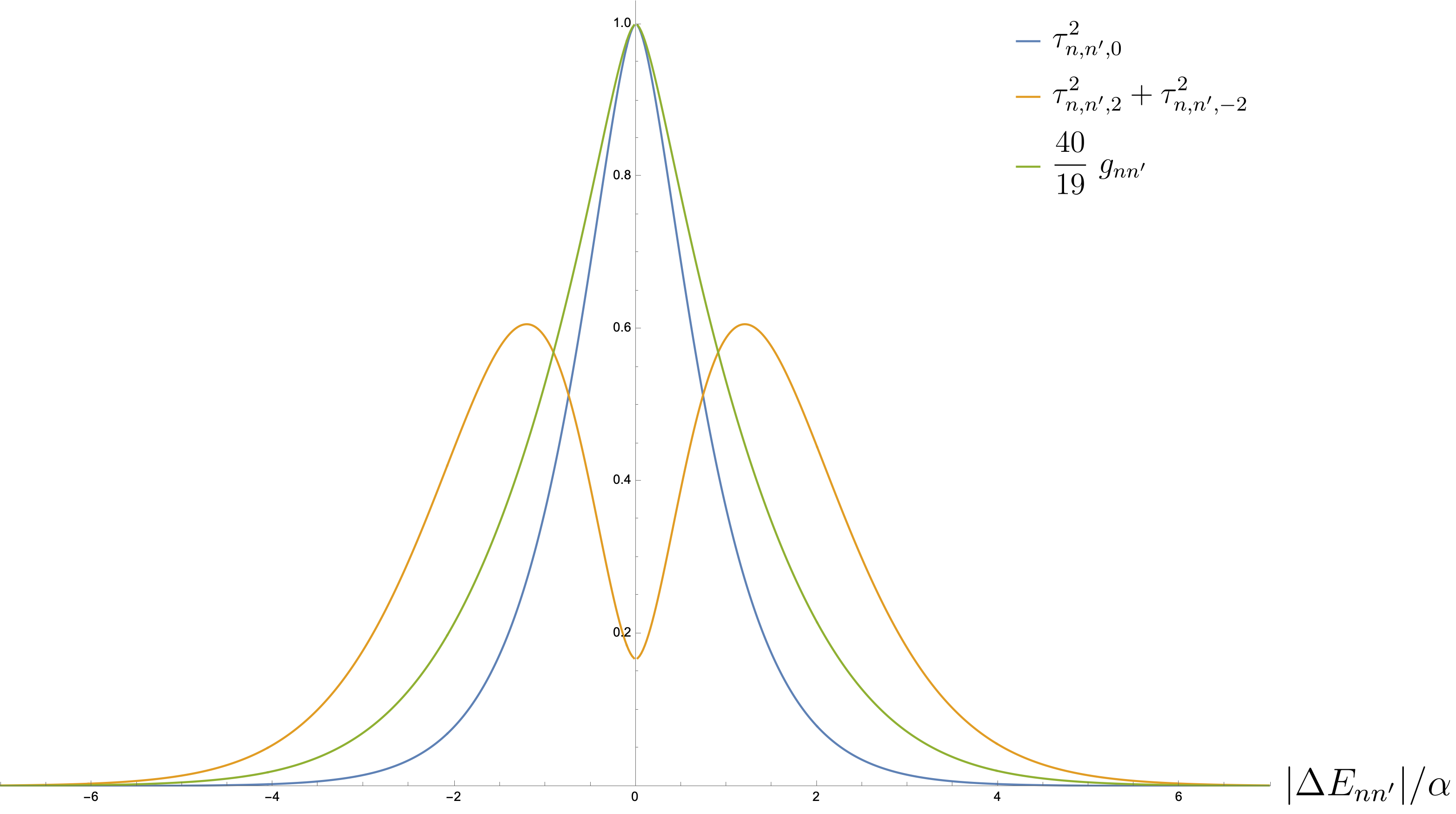}
 \caption{Window function entering the variation of the coefficients (green). The contributions from transitions with $\Delta l =0$ and $\Delta l =\pm 2$ are shown by the blue and orange lines, respectively.}
\label{fig:taus}
\end{figure}

Let's start by integrating Eq.~\eqref{eqci} with $i=1$.
 We choose a coordinate system with origin at the center of the clump and in which the trajectory of the star lies in the $x,y$ plane.
The star moves in the $y$ direction with constant velocity $v$ and it's the closest to the clump at $t=0$, when its distance to the center of the clump is $b$. In spherical coordinates, the position of the star is
\beq
\r_* = (r_*(t), \theta_*, \phi_*(t)) = (\sqrt{b^2+(vt)^2}, \pi/2, \arctan(vt/b))\enspace.
\eeq
We expand the Legendre polynomial appearing in $H_1$ as
\beq
P_2(\cos\gamma(t)) = \frac{4\pi}{5} \sum_{m=-2,0,2} Y^*_{2m}(\theta_*, \phi_*(t)) Y_{2m}(\theta, \phi) \enspace,
\eeq
where we used that $ Y^*_{2, \pm1}(\theta_*, \phi_*(t))=0$ for $\theta_*=\pi/2$.
We can break down the matrix element as follows
\beqa
\bra{nlm}H_1(t)\ket{n'l'm'} &=&  \frac{A}{\alpha} \bra{nl}r^2\ket{n'l'} \sum_{m''=0,\pm 2}I_{m''}^{lm,l'm'}T_{nn'm''}(t)\enspace,
\eeqa
where
\beqa
{\bra{nl}}r^2\ket{n'l'}
&=& \int dr\, r^4 R_{nl}(r)R_{n'l'}(r) \label{eq:ME}\\
I_{m''}^{lm,l'm'} &=& \sqrt{\frac{4\pi}{5}} \int d\Omega\ Y^*_{lm}(\theta, \phi)  Y_{l'm'}(\theta, \phi) Y_{2m''}(\theta, \phi)\\
T_{nn'm''}(t) &=&  \sqrt{\frac{4\pi}{5}} \frac{\alpha }{(1+(\alpha t)^2)^{3/2}}Y_{2m''}(\theta_*, \phi_*(t))\enspace.
\eeqa
Carrying out the integral in $I_{m''}^{lm,l'm'}$, we obtain the following selection rules
\beq
m''-m + m' =0 \qquad\qquad |l-2| \leq l' \leq l+2 \qquad\qquad l+l'=\mathrm{even}\enspace.
\eeq
In our setup, we have $l,l'\gg 2$. Moreover from the selection rules, we see that we can set $l'\approx l$ and $m'\approx m$. We will use these approximations everywhere except in the labels of the Winger functions (see below) and Kronecher deltas.
In the limit above, we obtain 
\beq
I_{m''}^{lm,l'm'} \approx \delta_{m', m-m''} (\delta_{l', l-2} + \delta_{l', l} + \delta_{l', l+2})\ d^2_{0,\, l - l'}(\pi/2)\ d^2_{m'',\, l - l'}(\arccos(m/l) )\enspace,
\eeq 
where $d^a_{b,c}(\theta)$ is a Wigner function.
We define
\beq
\tau_{nn'm} = \int_{-\infty}^\infty dt\ T_{nn'm''}(t)\ e^{i\Delta E_{nn'} t}\enspace,
\eeq
with $\Delta E_{nn'} = E_n-E_{n'}$. The functions $\tau_{nn'm''}$ can be evaluated analytically. However, we do not report the explicit expressions here, as they are not very enlightening. We can then integrate Eq.~\eqref{eqci} to obtain
\beqa
C^{(1)}_{nlm} &\approx& -i\frac{A}{\alpha} \sum_{n'}\ \sum_{l'=l,l\pm2}\ \ \sum_{m'=0,\pm 2} C^{(0)}_{n',l', m-m'}  \bra{nl}r^2\ket{n'l'}\  d^2_{0,\, l - l'}(\pi/2)\  \nl
&\times&
 d^2_{m',\, l - l'}(\arccos(m/l) )\ \tau_{n,n',m'}
 \enspace.
\eeqa
Now we calculate the ensemble average and evaluate the Wigner functions explicitly. This gives us the result
\beqa
\ev{|C^{(1)}_{nlm}|^2}_{ens} &\approx& \left(\frac{A}{\alpha}\right)^2
(4\pi)^2 m_a
\sum_{n',B}dE_{n'} \ \N_{n'l}\,f(E_{n'})\, \bra{nl}r^2\ket{n'l}^2 g_{nn'}^{lm}  \enspace,\label{C1sq_appr}
\eeqa
where
\beqa
g_{nn'}^{lm} =  \left( 1 - \frac{3m^2}{2l^2} \right)^2 \tau^2_{n,n',0} + \frac{3}{8}\frac{m^4}{l^4} (\tau^2_{n,n',2} + \tau^2_{n,n',-2}) \enspace.
\eeqa
The shapes of $\tau^2_{n,n',0}$ and $\tau^2_{n,n',2} + \tau^2_{n,n',-2}$ are shown in Fig.~\ref{fig:taus} as a function of $\Delta E_{nn'}/\alpha$. Notice that the sum in Eq.~\eqref{C1sq_appr}
runs over bound states only, as the unbound zeroth order coefficients are null.

A useful quantity for our computations will be average over $m$ of $\ev{|C^{(1)}_{nlm}|^2}_{ens}$ for fixed $n$ and $l$. We have in the large $l$ limit
\beqa
\ev{|C^{(1)}_{nlm}|^2}_{ens,m} 
&=&\frac{1}{2l+1}\sum_{m=-l}^l \ev{|C^{(1)}_{nlm}|^2}_{ens}\nl 
&\approx& \left(\frac{A}{\alpha}\right)^2
(4\pi)^2 m_a
\sum_{n'}dE_{n'} \ \N_{n'l}\,f(E_{n'})\, \bra{nl}r^2\ket{n'l}^2 g_{nn'} \enspace,\label{C1sq_appr_avgm}
\eeqa
where
\beqa
g_{nn'} &=&
\frac{1}{2l+1}\sum_{m=-l}^l g_{nn'}^{lm}\nl
&\approx&
\frac{3}{20}\left( 3 \tau^2_{n,n',0} +\frac{1}{2} (\tau^2_{n,n',2} + \tau^2_{n,n',-2})  \right)\enspace.
\eeqa
The function $g_{nn'}$ is shown in Fig.~\ref{fig:taus}. An excellent approximation is 
\beq
g_{nn'} \approx \frac{19}{40}\ e^{-\left(\frac{|\Delta E_{nn'}| }{c\alpha }\right)^a} \qquad\qquad a = 1.30415\qquad c = 1.42653 \enspace.
\label{eq:gnn}
\eeq

Let's now turn to the calculation of the second order coefficients. Integrating Eq.~\eqref{eqci} with $i=2$ and performing an ensemble average, we obtain
\beqa
\ev{{C_{nlm}^{(0)}}^* C_{nlm}^{(2)}+ c.c.}_{ens}
&=& -|C^{(0)}_{nlm}|^2\sum_{n'l'm'} 
\int_{-\infty}^\infty dt\int_{-\infty}^t dt'  \nl
&\times&\left( 
e^{i\Delta E_{nn'} (t-t')}\bra{nlm}H_1(\r,t)\ket{n'l'm'}
\  \bra{n'l'm'}H_1(\r,t')\ket{nlm} + c.c.\right)\nonumber\enspace.
\eeqa
We see that the integrand function is symmetric under $t\leftrightarrow t'$. We can then extend the integration $dt'$ to $\infty$ and divide the integrand by 2. The expression above reduces to
\beqa
\ev{{C_{nlm}^{(0)}}^* C_{nlm}^{(2)}+ c.c.}_{ens}
&=& -|C^{(0)}_{nlm}|^2\sum_{n'l'm'} 
\left|\int_{-\infty}^\infty dt\  
e^{i\Delta E_{nn'}t}\bra{nlm}H_1(\r,t)\ket{n'l'm'}\right|^2\nonumber\enspace.
\eeqa
Following the same steps as for the first order coefficients, we obtain
\beqa
\ev{{C_{nlm}^{(0)}}^* C_{nlm}^{(2)}+ c.c.}_{ens} &\approx& -\left(\frac{A}{\alpha}\right)^2
(4\pi)^2 m_a\ dE_{n} \ \N_{nl}\,f(E_{n})
\sum_{n'} \bra{nl}r^2\ket{n'l}^2 g_{nn'}^{lm}  \enspace,\nl\label{C2sq_appr}
\eeqa
where the sum over $n'$ runs over both bound and unbound states.

We can define the variation of the coefficients
\beqa
\Delta^{(+)} |C_{nlm}|^2 &=&
 \left(\frac{A}{\alpha}\right)^2
(4\pi)^2 m_a
\sum_{n'}dE_{n'} \ \N_{n'l}\,f(E_{n'})\, \bra{nl}r^2\ket{n'l}^2 g_{nn'}\nl
&=&
 \left(\frac{A}{\alpha}\right)^2
\sum_{n'}|C^{(0)}_{n'lm}|^2\, \bra{nl}r^2\ket{n'l}^2 g_{nn'} \\
\Delta^{(-)} |C_{nlm}|^2 &=&
-\left(\frac{A}{\alpha}\right)^2
(4\pi)^2 m_a\ dE_{n} \ \N_{nl}\,f(E_{n})
\sum_{n'} \bra{nl}r^2\ket{n'l}^2 g_{nn'}\nl
&=&
-\left(\frac{A}{\alpha}\right)^2
|C^{(0)}_{nlm}|^2\
\sum_{n'} \bra{nl}r^2\ket{n'l}^2 g_{nn'} \enspace.
\eeqa

\subsection{On the impulse approximation}
\label{sec:impulsapprox}

Let us consider the combination $|\bra{nl}r^2\ket{n'l}|^2 g_{nn'} \equiv |\mathcal{M}|^2 g_{nn'}$ appearing in
Eq.~\eqref{eq:DE1}. Both factors can be considered as functions of $\epsilon \equiv \Delta E_{nn'} = E_n - E_{n'}$, which are peaked around $\epsilon = 0$ and have a characteristic width $\sigma_\epsilon$, such that values for $\epsilon \gg \sigma_\epsilon$ are suppressed. In the following we show that $\sigma_\epsilon^\mathcal{M} \ll \sigma_\epsilon^g$. In this case, $g_{nn'}$ can be considered as constant, $g_{nn'}(\epsilon) \approx g_{nn'}(0) \approx 1/2$ (see Eq.~\eqref{eq:gnn})
and pulled out of the integral over $E'$.

From Eq.~\eqref{eq:gnn} we see that $\sigma_\epsilon^g \simeq \alpha = v/b$. The width of $|\mathcal{M}|^2$ can be estimated from Eq.~\eqref{eq:ME}. It emerges from the averaging of the oscillating terms in the radial wave functions when integrated over $r$. The relevant phase difference is of order
\begin{align}
    \Delta\varphi &= [\sqrt{2m_a(E - V_l)} - \sqrt{2m_a(E' - V_l)}] [r_2(E,l) - r_1(E,l)] \nonumber\\
    &\simeq \epsilon \sqrt{\frac{m_a}{2(E - V_l)}} [r_2(E,l) - r_1(E,l)] 
    \sim \epsilon \sqrt{\frac{R}{GM}} R \,, 
\end{align}
where $r_1, r_2$ are the classical turning points. For $\Delta\varphi \gg 1$ the oscillating terms will average to zero, and therefore we have $\sigma_\epsilon^\mathcal{M} \sim \sqrt{GM/R^3}$. Hence we obtain
\begin{align}
    \frac{\sigma_\epsilon^\mathcal{M}}{\sigma_\epsilon^g} &\sim
    \frac{b}{R} \, \frac{1}{v} \sqrt{\frac{GM}{R}} \nonumber\\
    &\simeq 2\times 10^{-7} \frac{b}{R} 
    \left(\frac{10^{-3}}{v}\right)
    \left(\frac{M}{10^{-12}M_\odot}\right)^{1/2}
    \left(\frac{10^{-6}\, {\rm pc}}{R}\right)^{1/2} \,,
\end{align}
which shows that for typical parameters the approximation 
$\sigma_\epsilon^\mathcal{M} \ll \sigma_\epsilon^g$ is well justified. 

Note that $\sqrt{GM/R}$ corresponds to a typical velocity $w$ of particles bound to the cluster. Hence, we recover the impulse approximation in the particle picture, which means that the time scale $b/v$ for the duration of the star encounter is much shorter than the time $R/w$ needed for a bound particle to cross the cluster, i.e., particles in the cluster can be considered ``at rest'' during the encounter, see e.g., \cite{Green:2006hh}.

\section{Stellar and dark matter distributions in the Milky Way }
\label{app:distribution}

\subsection{Stellar population}
\label{sec:StellarPopulation}

\begin{figure}[t]
\centering
  \includegraphics[scale=0.36]{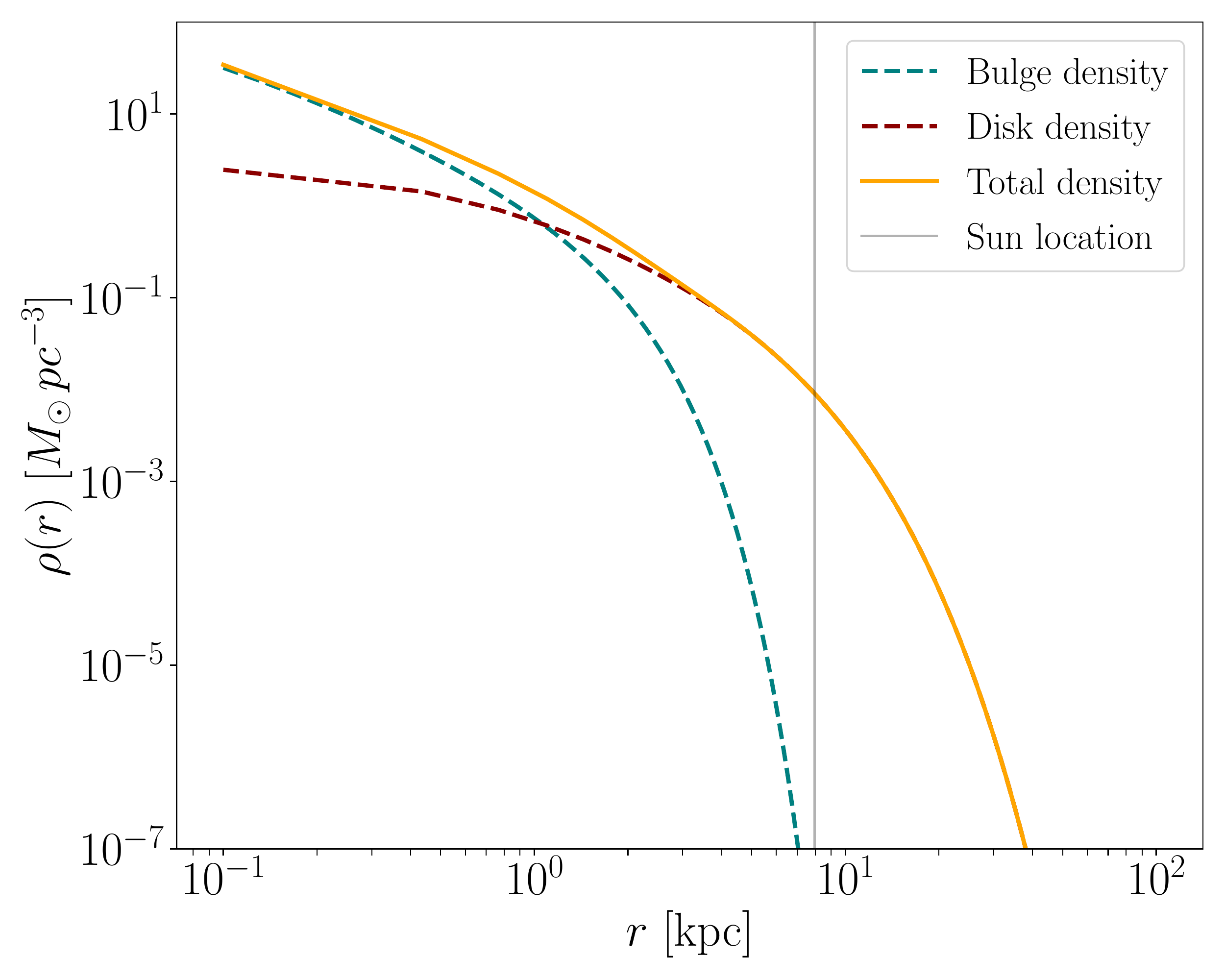}
 \caption{Spherically averaged stellar density distributions in the Milky Way for bulge and disk separated (dashed) and the total density (solid). The sun location is indicated by the gray vertical line at $r_\odot \approx 8$ Kpc.}
\label{fig:Stellar Distribution}
\end{figure}

The star distribution in the galaxy can be decomposed into a high-density and quickly decaying central bulge and a more extended disk. The bulge is expected to have a triaxial shape \cite{Portail_2015}; for simplicity, we approximate it by an axisymmetric profile \cite{McMillan_2011}
\begin{equation}
   \rho^{\text{bulge}}(r_c,z) =\rho_0 \frac{\text{e}^{-(r'/r_{\text{cut}})^2}}{\left(1+r'/r_{0}\right)^{\lambda}},
\end{equation}
with $r_{\text{cut}} = 2.1$ kpc, $\lambda=1.8$, $r_0= 0.075$ kpc, $\rho_0 = 99.3 \,M_\odot \, \text{pc}^{-3}$, $r' =\sqrt{r_c^2+\left(2z\right)^2}$, and $r_c$ denotes the radial coordinate in cylindrical coordinates. We  average over the angles to obtain a spherically symmetric distribution. Note, however, that different models exist, leading to more or less similar expression (see for instance Ref.~\cite{Misiriotis:2006qq}).

The disk is generally composed of two sub-disks, thin (t) and thick (T) disks, each of them described by an axisymmetric double exponential model \cite{Dehnen:1996fa}:
\begin{equation}
    \rho^{\text{disk}}(r_c,z) = \frac{\Sigma_d}{2 z_d}\text{exp}\bigg(-\frac{r_c}{r_d}-\frac{|z|}{z_d}\bigg),
\end{equation}
with $\Sigma_d^t = 816.6 \,M_\odot \, \text{pc}^{-2}$, $\Sigma_d^T = 209.5 \,M_\odot \, \text{pc}^{-2}$, $r_d^t = 2.90\, \text{kpc}$, $r_d^T = 3.31\, \text{kpc}$, $z_d^t = 0.3\, \text{kpc}$, $z_d^T = 0.9\,\text{kpc}$ \cite{Kavanagh:2020gcy}. Again, we average over angles to have a spherically symmetric profile.

The sum of the bulge and the disk densities is shown in Fig.~\ref{fig:Stellar Distribution}.
We use this distribution to calculate the total number of interactions a given minicluster undergoes. In particular, assuming that the stars all have the same mass of $1 \, M_\odot$, their number density is given by
\begin{equation}
    n_*(r) = \frac{1}{1\, M_\odot} \left( \rho^{\text{disk}}(r) +  \rho^{\text{bulge}}(r)\right) \,.
\end{equation}

\subsection{NFW profile from axion miniclusters}
\label{sec:NFWreconstruction}

We assume that the dark matter halo of the Milky Way is initially composed of axion miniclusters only and has an NFW density profile~\cite{Navarro:1996gj}
\begin{equation}
  \rho_{\text{NFW}}(r) = \frac{\rho_s}{\left(r/r_s\right)\left(1+r/r_s\right)^2} \,,
  \label{NFW profile}
\end{equation}
with the parameters $\rho_s=0.014 \, M_\odot \, \text{pc}^{-3}$ and  $r_s=16.1 \, \text{kpc}$~\cite{Nesti:2013uwa}, obtained by fitting the rotation curve extracted from atomic hydrogen measurement~\cite{McClure-Griffiths:2007ddt}.
Our aim is now to derive the local probability distributions for the orbital parameters for the miniclusters that reproduce this profile. In particular, each minicluster orbit will be characterized by its semi-major axis $a$ and eccentricity $e$.
Our approach is similar to the one of Ref.~\cite{Kavanagh:2020gcy}.

It can be shown that the NFW profile $\rho_{\text{NWF}}(r)$ can be reconstructed by using a specific probability distribution $P(e)$ for the eccentricity and by assuming that the density as a function of the semi-major axis follows the NFW profile, i.e $\rho(a) \approx \rho_{\text{NWF}}(a)$. Hence, 
\begin{equation}
    \rho_{\text{NFW}}(r) \approx \int da\, 4\pi a^2 \rho_{\text{NFW}}(a) \int de\,P(e) \frac{P(r|a,e)}{4\pi r^2}, 
    \label{NFW reconstruction}
\end{equation}
with the eccentricity probability distribution \cite{vandenBosch:1998dc}
\begin{equation}\label{eq:Pe}
    P(e) = 4e(1-e^2) \,,
\end{equation}
and the conditional probability of being at radius $r$ for given orbital parameters $(a,e)$
\begin{equation}
    P(r|a,e)=\frac{2}{T_{\text{orb}}}\left(\frac{dr}{dt}\right)^{-1} \,,
    \label{Conditional Proba}
\end{equation}
where
\begin{equation}
    \begin{split}
        &\frac{dr}{dt} =\sqrt{GM_G(a)\left(  \frac{2}{r}-\frac{1}{a}-\frac{a(1-e^2)}{r^2}\right)} \,,\\
    &T_{\text{orb}}=2\pi\sqrt{\frac{a^3}{GM_G(a)}} \,,
\label{OrbitPeriod}
    \end{split}
\end{equation}
with $M_G(a)$ being the mass enclosed inside a sphere of radius $a$, 
and $a(1-e)\leq r \leq a(1+e)$.

Our goal is now to obtain $P(a,e|r)$. We first use that $P(e)$, Eq.~\eqref{eq:Pe}, is independent of $r$ and $a$, and therefore $P(a,e|r) = P(a|r,e)P(e)$. Next we use Bayes theorem:
\begin{equation}
\begin{split}
    P(r,a,e) &= P(r|a,e) P(a) P(e) \\
             &= P(a|r,e) P(r) P(e) \,.
\end{split}
\end{equation}
Since the density profile as a function of the radius or the semi-major axis is given by the same function (namely $\rho_{\rm NFW}$, see above) we can assume that $P(r) \approx P(a)$. It follows that $P(r|a,e) = P(a|r,e)$ and hence, $P(a|r,e)$ is directly given by Eq.~\eqref{Conditional Proba}.

In conclusion, at a given radius $r$, the distribution of the minicluster orbital parameters is given by $P(e)$ in eccentricity and by $P(a|r,e)$ in semi-major axis. We have checked by randomly drawing $e$ and $a$ according to the procedure outlined above, that the initial NFW profile $\rho_{\rm NFW}(r)$ is recovered to good accuracy. Note that these distributions are independent of the mass and radius distributions of the miniclusters.


\bibliographystyle{JHEP_improved}
\bibliography{./refs}

\end{document}